\documentclass[10pt,journal,compsoc]{IEEEtran}

\usepackage[justification=centering]{caption}
\usepackage{algorithm,algpseudocode}
\usepackage{amsmath}
\usepackage{graphicx}
\usepackage{times}                               
\usepackage{graphicx}
\usepackage{tabularx}
\usepackage{array}
\usepackage{amsmath}
\usepackage{amssymb}
\usepackage{units}
\usepackage{textcomp,xspace}
\usepackage{enumerate}
\usepackage{multirow}
\usepackage{multicol}
\usepackage{subfigure}
\usepackage{color}

\makeatletter
\def\BState{\State\hskip-\ALG@thistlm}
\makeatother

\newcommand{\R}{\mathbb{R}}

\ifCLASSINFOpdf

\else

\fi

\begin{document}

\title{Accelerating BLAS on Custom Architecture through Algorithm-Architecture Co-design}

\author{Farhad Merchant,
        Tarun Vatwani, Anupam Chattopadhyay, ~\IEEEmembership{Senior Member, IEEE,} Soumyendu Raha, \\S K Nandy, ~\IEEEmembership{Senior Member, IEEE,} and Ranjani Narayan
\IEEEcompsocitemizethanks{\IEEEcompsocthanksitem Farhad Merchant and Anupam Chattopadhyay are with School of Computer Science and Engineering,
Nanyang Technological University, Singapore\protect\\
E-mail: \{mamirali,anupam\}@ntu.edu.sg
\IEEEcompsocthanksitem Tarun Vatwani is with Indian Institute of Technology, Jodhpur 
\IEEEcompsocthanksitem Soumyendu Raha and S K nandy are with Indian Institute of Science, Bangalore \IEEEcompsocthanksitem Ranjani Narayan is with Morphing Machines Pvt. LTd. }
\thanks{Manuscript received October 20, 2016;}}

\IEEEtitleabstractindextext{%
\begin{abstract}
Basic Linear Algebra Subprograms (BLAS) play key role in high performance and scientific computing applications. Experimentally, yesteryear multicore and General Purpose Graphics Processing Units (GPGPUs) are capable of achieving up to 15 to 57\% of the theoretical peak performance at 65W to 240W respectively for compute bound operations like Double/Single Precision General Matrix Multiplication (XGEMM). For bandwidth bound operations like Single/Double precision Matrix-vector Multiplication (XGEMV) the performance is merely 5 to 7\% of the theoretical peak performance in multicores and GPGPUs respectively. Achieving performance in BLAS requires moving away from conventional wisdom and evolving towards customized accelerator tailored for BLAS through algorithm-architecture co-design. In this paper, we present acceleration of Level-1 (vector operations), Level-2 (matrix-vector operations), and Level-3 (matrix-matrix operations) BLAS through algorithm architecture co-design on a Coarse-grained Reconfigurable Architecture (CGRA). We choose REDEFINE CGRA as a platform for our experiments since REDEFINE can be adapted to support domain of interest through tailor-made Custom Function Units (CFUs). For efficient sequential realization of BLAS, we present design of a Processing Element (PE) and perform micro-architectural enhancements in the PE to achieve up-to 74\% of the theoretical peak performance of PE in DGEMM, 40\% in DGEMV and 20\% in double precision inner product (DDOT). We attach this PE to REDEFINE CGRA as a CFU and show the scalability of our solution. Finally, we show performance improvement of 3-140x in PE over commercially available Intel micro-architectures, ClearSpeed CSX700, FPGA, and Nvidia GPGPUs. 
\end{abstract}

}

\maketitle

\IEEEpeerreviewmaketitle

\section{Introduction}
\label{sec:intro}

\begin{figure}[!h]
	\centering
	\includegraphics[scale=0.20]{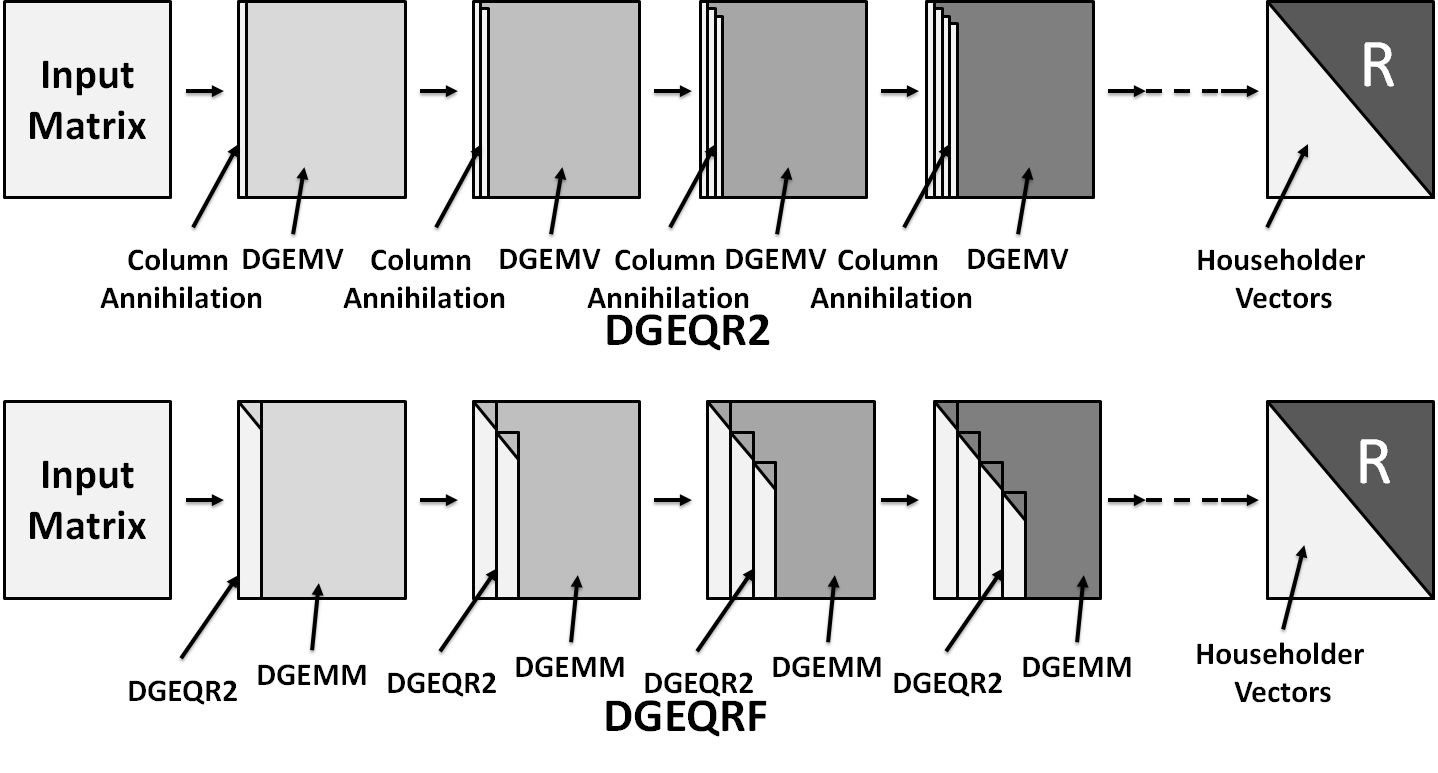}
	\caption{DGEQR2 and DGEQRF Routines}
	\label{fig:dgeqr2_dgeqrf}
\end{figure}

Several engineering and scientific applications require solution of dense linear systems of equations and linear least square problems where matrix factorizations like LU, QR and Cholesky methods play pivotal role. Traditionally, routines of these factorizations that are part of Linear Algebra Package (LAPACK) are written as a series of Basic Linear Algebra Subprogram (BLAS) calls \cite{laug}\cite{lapack1}. Pictorial representation of double precision QR factorization routines, DGEQR2 and DGEQRF that are part of LAPACK is shown in the figure \ref{fig:dgeqr2_dgeqrf}. 

In the pictorial representation of DGEQR2 it can be observed that DGEQR2 is dominated by matrix-vector operations (DGEMV in BLAS) and DGEQRF is dominated by DGEQR2 and matrix-matrix operations (DGEMM in BLAS). Our experiments for DGEQR2 on Intel Core i7 and observation using Intel VTune\texttrademark$\;$suggests that for matrix of size $10k\times 10k$, 99\% of the total time DGEMV executes while double precsion inner product (DDOT) executes for hardly 1\% of the total time in the operation of DGEQR2. Similarly, DGEQRF is dominated by DGEMM and it runs for 99\% of the total time of DGEQRF while DGEQR2 runs for 1\% of the time. Similar observations can be made in the routines like XGETRF (double/single precision LU factorization routine) and XPBTRF (double/single precision Cholesky factorization routine). Considering importance of BLAS in LAPACK, it is arguably one of the most interesting research problem to accelerate BLAS. 

For acceleration of BLAS, a library based approach is adopted. Based on reference BLAS and LAPACK available on Netlib, Intel Math Kernel Library (MKL), IBM's Engineering and Scientific Subroutine Library (ESSL), AMD's AMD Core Math Library (ACML), Nvidia's CUDA Linear Algebra (CULA) where CULA dense is for Dense Linear Algebra (DLA) and CULA sparse is for Sparse Linear Algebra (SLA), and cuBLAS which is yet another CUDA Baic Linear Algebra Subprograms are developed. There are also several open source packages for multores and General Purpose Graphics Processing Units (GPGPUs) based realizations like Parallel Linear Algebra Software for multicore Architectures (PLASMA) and Matrix Algebra on Multicore and GPGPU Architectures (MAGMA) use BLAS as a basic building block. All these mentioned packages are developed for multicore and GPGPUs for realization of DLA computations in the most efficient way. PLASMA and MAGMA incorporate tiled algorithms that are capable of exploiting memory hierarchy efficiently \cite{plasma1}\cite{magma1}. Despite all the efforts being directed towards acceleration of DLA computations, the performance attained by yesteryear platforms is as low as 15-17\% of the theoretical peak performance in multicore and 55-57\% of the theoretical peak performance in GPGPU at $\ge$65W and $\ge$240W power consumption respectively. Considering inability of GPGPU and multicore architectures in exploiting parallelism available in BLAS, we recommend algorithm-architecture co-design for BLAS as a solution for efficient realization of DLA. Performance of several recent realizations in detail is discussed in section \ref{sec:rw}.



Recently, Coarse-grained Reconfigurable Architectures (CGRAs) have gained popularity due to their power performance and flexibility \cite{drra1}\cite{adres2}\cite{cgr1}. Performance advantage in CGRAs is attained by supporting selected number of data-paths out of all possible data-paths and hence they occupy middle ground between Application Specific Integrated Circuits (ASICs) and Field Programmable Gate Arrays (FPGAs) \cite{chiou1}\cite{egra1}\cite{cgr2}\cite{cgr3}. CGRAs like REDEFINE have special feature that they can be customized for application domains where several data-paths belonging to a particular domain of interest are realized as a reconfigurable ASIC \cite{Alle1}. In REDEFINE, several Tiles are connected through a Network-on-Chip (NoC) where Custom Function Units (CFUs) tailored for a particular domain decides performance of overall system for application domain \cite{Alle1}\cite{noc2}. REDEFINE is shown in figure \ref{fig:mm_2x2} along with several Tiles in a simulation environment.

Major contributions in this paper are as follows:
\begin{itemize}
	\item Firstly, we present evaluation of legacy BLAS on off-the-shelf Intel/AMD processor and Nvidia GPGPUs where Cycles-per-Instruction (CPI) and Gflops/watt (Giga flops per watt) based analysis  is discussed. Through detailed experiments it is shown that with the best efforts the performance achieved for BLAS in Intel/AMD and Nvidia GPGPU is between 0.02 to 0.25 Gflops/watt
	\item We present Directed Acyclic Graph (DAG) based analysis of representative routines of Level-1, Level-2, and Level-3 BLAS and discuss available parallelism and possible data locality in these routines. We also identify macro operations and realize them on a Reconfigurable Data-path (RDP). Based on our analysis, we arrive at design of a Processing Element (PE)
	\item Several architectural enhancements are performed in the PE presented in \cite{Merc1} for improving throughput of BLAS by exploiting parallelism and data locality. These enhancements result in efficient realization of sequential BLAS in the PE. In this exposition, we have extended scope of experiments to accommodate matrix sizes of $80\times 80$ and $100\times 100$ to bring more clarity of saturation in the attained performance after each enhancement 
	\item It is shown that through algorithm-architecture co-design, we are able to break the saturation point with each enhancement and improve the overall performance of the BLAS in PE. With each enhancement, we show that we are able to push the saturation point towards theoretical peak performance of the PE at very high energy efficiency
	\item We attach the PE to the Routers in REDEFINE for parallel realization of BLAS and show algorithmic and architecture scalability
\end{itemize}   

The organization of the paper is as follows: In section \ref{sec:rw}, some of the recent Multi-core, GPGPU, and custom realizations of BLAS are discussed. In section \ref{sec:mot}, we present legacy BLAS realization on multicore and GPGPU and CPI and energy efficiency analysis of the realization. In section \ref{sec:graph}, DAG based analysis of Level-1, Level-2, and Level-3 BLAS is presented and we derive preliminary specifications of a PE. Architectural enhancements in the PE for improvement in throughput in BLAS and parallel realization of BLAS on REDEFINE where we attach PE as a CFU to REDEFINE are presented in section \ref{sec:real} and the work is summarized in section \ref{sec:con}.
\section{Related Work}
\label{sec:rw}
Over the years there have been several efficient realization of BLAS due to applicability in high performance scientific application. In this section, we survey several multicore and GPU based, and custom realizations of BLAS. We consider FPGA based realizations as custom realizations. 

\subsection{Software Packages for Multicore Platforms}

The first ever software library LINPACK for performing linear algebra computations was developed in 1970s and early 1980s \cite{linpack1}. Subsequently, LINPACK that used Level-1 BLAS as a basic building block was superseded by LAPACK that uses Level-3 BLAS as a basic building block \cite{lapack1}. In the recent years, with arrival of multicore architectures, there have been several advancements in the parallel realization of LAPACK. One such effort is PLASMA, that can perform computations on multicore architecture with the help of dynamic scheduler Queuing and Runtime for Kernels (QUARK). PLASMA creates pipeline model for parallel execution by dynamic scheduling of BLAS kernels on the multicore platform \cite{quark1}\cite{plasma1}. A Formal Linear Algebra Method Environment (FLAME) focuses on issues related to programming of linear algebra programs. The focus of the FLAME project is to automatically generate efficient linear algebra codes for the underlying platform \cite{flame3}\cite{flame1}. Under the umbrella of FLAME project, BLAS-like Library Instantiation Software (BLIS) focuses on rapid scheduling of BLAS-like kernels on multicore architectures. Automatically Tuned Linear Algebra Software (ATLAS) is an matured open source package that generates BLAS for the underlying platform \cite{atlas1}\cite{atlas2}. ATLAS relies on legacy BLAS for generation of efficient code for the underlying platform where several parameters are tweaked to suit the underlying platform. OpenBLAS is another open source package that focuses on efficient realization of DLA computations \cite{openblas1}\cite{openblas2}. OpenBLAS relies on GotoBLAS for the performance where GotoBLAS is a set of assembly programs written for DLA computations. A major shortcoming of the packages like LAPACK, PLASMA, BLIS, ATLAS, and OpenBLAS is lack of support from the underlying platform resulting in 15-20\% of the theoretical peak performance. 

\subsection{GPU Based Realizations}
GPUs were originally designed for graphics processing are highly suitable for general purpose computing. There have been several packages developed to perform efficient BLAS functionality on GPUs. The most prominent of all of them is MAGMA software package \cite{magma1}. MAGMA relies on MAGMA BLAS for the performance where the performance of MAGMA DGEMM is observed to be 57\% of the peak performance of Tesla C2050 GPU with theoretical peak of 512 Gflops for double precision. KAUST BLAS (KBLAS) is one of the most recent and ongoing research project at KAUST. KBLAS internally relies on Cuda BLAS developed by Nvidia for the performance on GPU. BLASX presented in \cite{blasx1} focuses on optimizaiton of Level-3 BLAS in multi-GPU environment. BLASX minimizes the global communication through two level hierarchical tile cache structures and achieves 92.68\% of the in-core cuBLAS DGEMM. BLASX also contains better load balancing techniques compared to MAGMA and cuBLAS. Despite elegant scheduling technique and efficient exploitation of memory hierarchy BLASX, the performance achieved by BLASX is limited by cuBLAS DGEMM. In \cite{superlinear1}, requirement for cache memory is studied in detail for achieving superliear speed-up for XGEMM. The study presented in \cite{superlinear1} has no mention of data type if it is single precision or double precision. All the GPU based realization of BLAS fail to achieve high performance due to lack of support for GEMM primitives.   

\subsection{Custom Realizations}
Customized accelerators are the class of architectures that are tuned for low energy, and unit area at high throughput for domain of interest \cite{exp3}\cite{lac3}. Cell Broadband Engine (CBE) from International Business Machine (IBM) is a high performance architecture designed based on Power PC core \cite{cellbe1}. Due to energy efficiency of CBE, it is viewed as an ideal platform for scientific computing \cite{cellbe2}. 
ClearSpeed's CSX architecture is back bone of ClearSpeed CSX600 and CSX700 processors. These processors have very high energy efficiency and operate at 12 Watts with theoretical peak of 96 GFlops \cite{clearspeed1}\cite{clearspeed2}\cite{lac1}. A major shortcoming of ClearSpeed's CSX and CBE architectures are low Gflops/W and Gflops/$mm^2$ \cite{lac1}. 

There have been several attempts in viewing Field Programmable Gate Arrays (FPGAs) as high performance computing engines \cite{lapack_fpga1}\cite{lapack_fpga2}\cite{blas_fpga1}. Mostly FPGAs are envisioned as a high performance co-processor of a programmable host processor for compute intensive applications \cite{blas_fpga2}\cite{blas_fpga3}. A major disadvantage of FPGAs is higher power consumption than an Application Specific Integrated Circuit (ASIC) counterpart of the same logic. FPGAs are also limited by the on-chip logic resulting in scalability issues in high performance computing applications. 

To overcome shortcomings of the existing architectures in exploiting parallelisms available in DLA computations, we take a route of algorithm-architecture co-design where we ensure high performance along with energy, area efficiency, and scalability.

\section{BLAS Realization on Off-the-shelf Processors}\label{sec:mot}

GEMM (Level-3 BLAS) and GEMV (Level-2 BLAS) are the most prominent routines in many engineering and scientific computations. These routines also have pedagogical importance due to its simplistic nature and often used to evaluate emerging architectures. In this section, first we discuss GEMM algorithm and then we examine some of the recent realization of GEMM. Based on the anslysis of the profiling of the BLAS routines, we arrive at the pitfalls in extracting performance out of GEMM on contemporary multicore and GPU platforms. We further decide to design our own customized platform that is capable of extracting performance in BLAS through efficiently exploiting parallelism in BLAS.  

\subsection{GEMM and GEMV Algorithms}

\begin{algorithm}
\caption{GEMM - General Matrix Multiplication}
\label{algo:gmm1}
\begin{algorithmic}[1]
\State Allocate memories for input and output matrices and initialize input matrices
\For{$i=1$ to $m$}
  \For{$j=1$ to $n$}
    \For{$k=1$ to $n$}
      \State C(i,j) = A(i,k)B(k,j) + C(i,j)
     \EndFor
  \EndFor
\EndFor
\end{algorithmic}
\end{algorithm}

\begin{algorithm}
\caption{GEMV - Matrix Vector Multiplication}
\label{algo:gmv1}
\begin{algorithmic}[1]
\State Allocate memories for input matrix, input vector and output vector. Initialize input matrix and input vector
  \For{$j=1$ to $n$}
    \For{$i=1$ to $m$}
      \State y(i) = A(i,j)x(j) + y(i)
     \EndFor
  \EndFor
\end{algorithmic}
\end{algorithm}


Pseudo codes for GEMM and GEMV are described in algorithms \ref{algo:gmm1} and \ref{algo:gmv1} respectively. GEMM algorithm has three loops and hence it belongs to Level-3 BLAS while GEMV belongs to Level-2 BLAS. For multiplying two matrices of size $n\times n$, it takes $n^3$ multiplications and $n^3-n^2$ additions while GEMV takes $n^2$ multiplications and $n^2 - n$ additions. Typically, GEMM and GENV exhibit Instruction Level Parallelism (ILP) and Data Level Parallelism (DLP). GEMM also exhibits mighty data locality and is capable of sustaining $O(n)$ computations to communication ratio. All together, if exploited efficiently and accelerated, GEMM and GEMV become an ideal candidate to be used as a basic building block for many high performance scientific applications. Since, GEMM has three nested loops, these loops can be permuted to change the access pattern of input matrices as shown in table \ref{tab:gmm_tab}. 

\begin{table}[!h]
\caption{General Matrix Multiplication (GEMM): Loop Orderings and Access Patterns\label{tab:gmm_tab}}{
\centering
\begin{tabular}{ |p{0.8cm}|p{1cm}|p{2.5cm}|c| } 
 \hline
 Loop Order & Inner Loop & Middle Loop & Inner Loop Data Access \\  \hline \hline
ijk & $dot$ & $vector\times matrix$ & A by row, B by column \\ \hline 
jik & $dot$ & $matrix\times vector$ & A by row, B by column \\ \hline
ikj & $saxpy$ & row $gaxpy$ & B by row, C by row \\ \hline
jki & $saxpy$ & column $gaxpy$ & A by column, C by column \\ \hline
kij & $saxpy$ & row outer product & B by row, C by row \\ \hline
kji & $saxpy$ & column outer product & A by column, B by column \\ \hline
\end{tabular}}
\end{table}

In the table \ref{tab:gmm_tab}, $saxpy$ stands for "scalar $a$ multiplied by vector $x$ plus vector $y$" and $gaxpy$ stands for generalized $saxpy$ \cite{Golub1}\cite{Higham_book}. Further details of GEMM can be found in \cite{Golub1}, and \cite{Higham_book}.

\subsection{Performance Evaluation of GEMM and GEMV}
For contemporary architecture, highly efficient GEMM is realized as a subroutine in BLAS. There exists several vendor specific realizations of DGEMM. For our experiments, we take DGEMM available in BLAS from The Netlib and for evaluation on GPU we use MAGMA\_DGEMM. We compile DGEMM for different Intel and AMD machines with different compiler options and evaluate the performance of DGEMM for these architectures. We evaluate MAGMA\_DGEMM on Telsa C2050.

\begin{figure*}[]
\centering
\subfigure[CPI in DGEMM on Intel Haswell and AMD Bulldozer Micro-architectures\label{fig:cpi_gmm_0}]{\includegraphics[scale = 0.18]{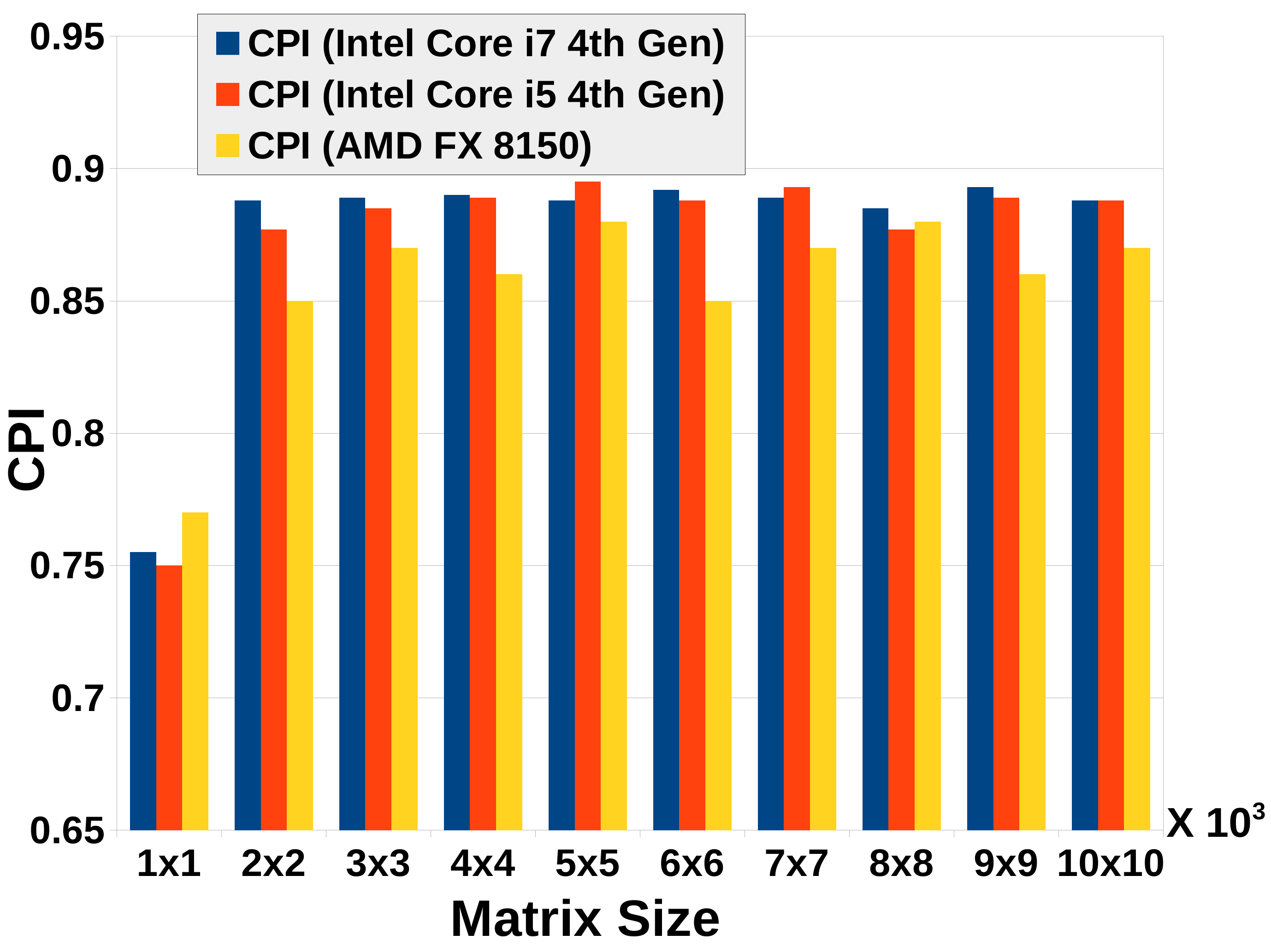}}
\subfigure[Gflops in DGEMM on Intel Haswell and AMD Bulldozer Micro-architectures\label{fig:cpi_gmm_1}]{\includegraphics[scale = 0.18]{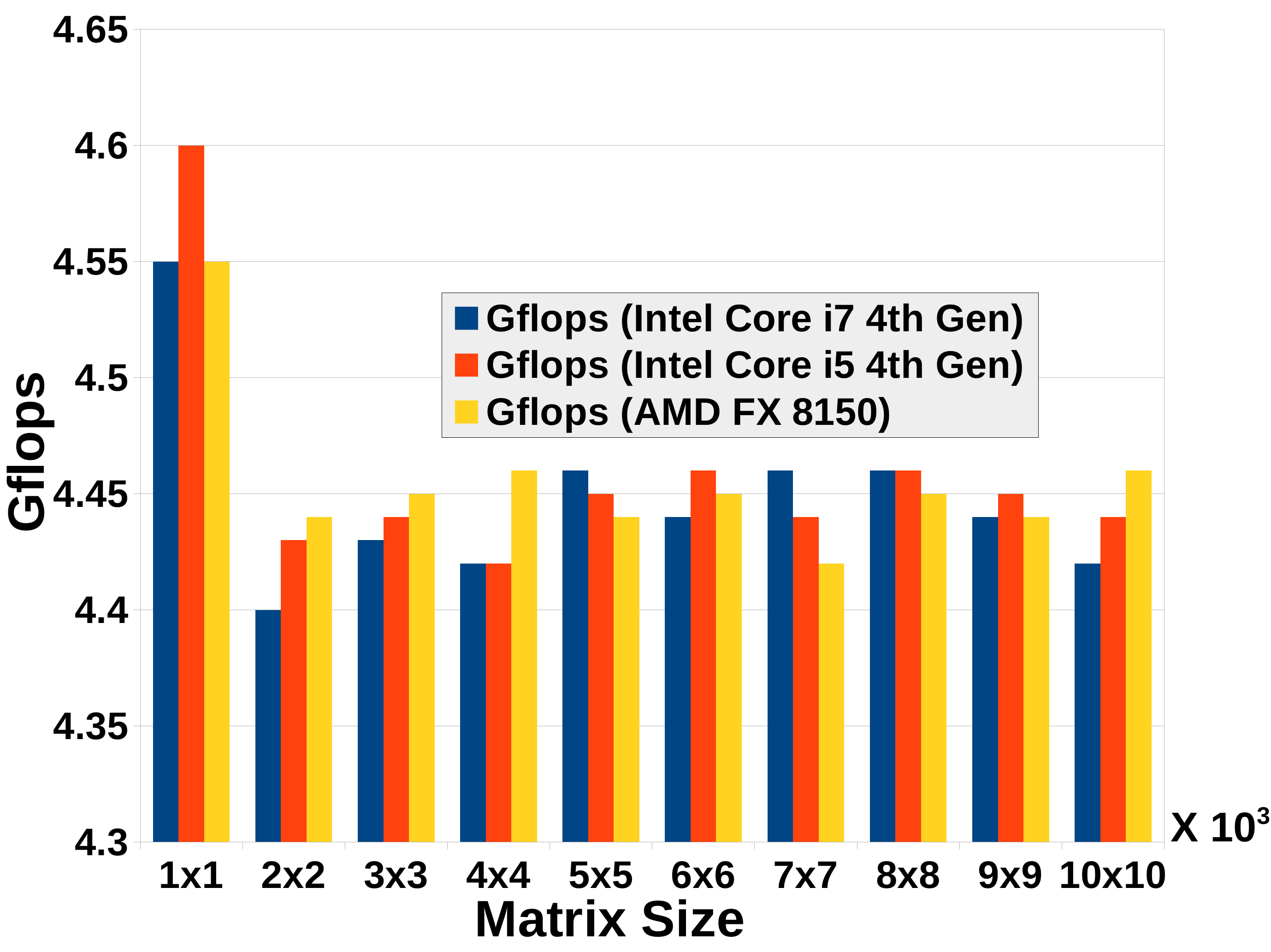}}
\subfigure[CPI in DGEMM on Intel Haswell Micro-architecture with $icc$\label{fig:cpi_gmm_2}]{\includegraphics[scale = 0.18]{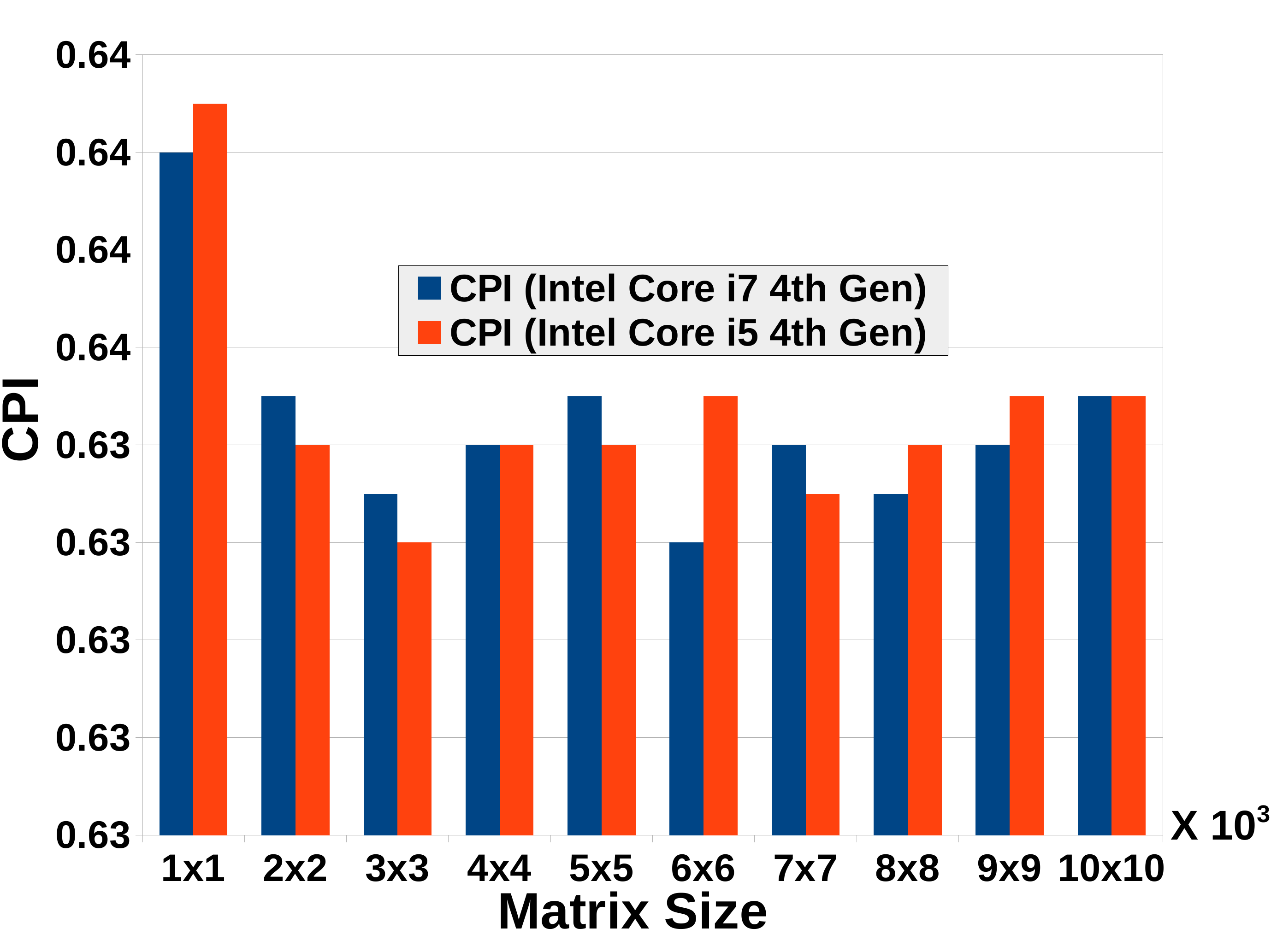}}
\subfigure[Gflops in DGEMM on Intel Haswell Micro-architecture with $icc$\label{fig:cpi_gmm_3}]{\includegraphics[scale = 0.18]{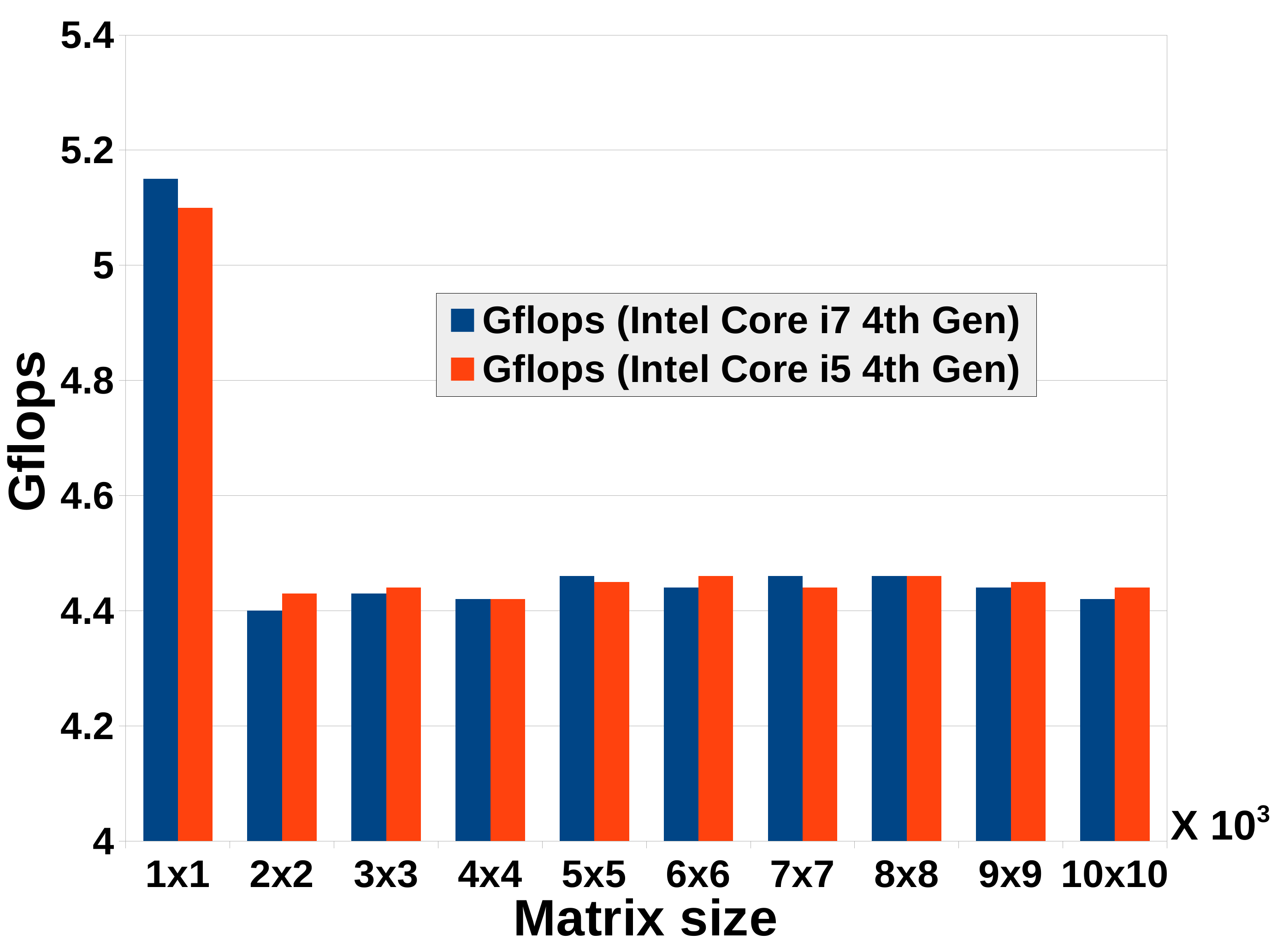}}
\subfigure[CPI in DGEMM on Intel Haswell Micro-architecture with $icc$ and $-mavx$ compiler switches\label{fig:cpi_gmm_4}]{\includegraphics[scale = 0.18]{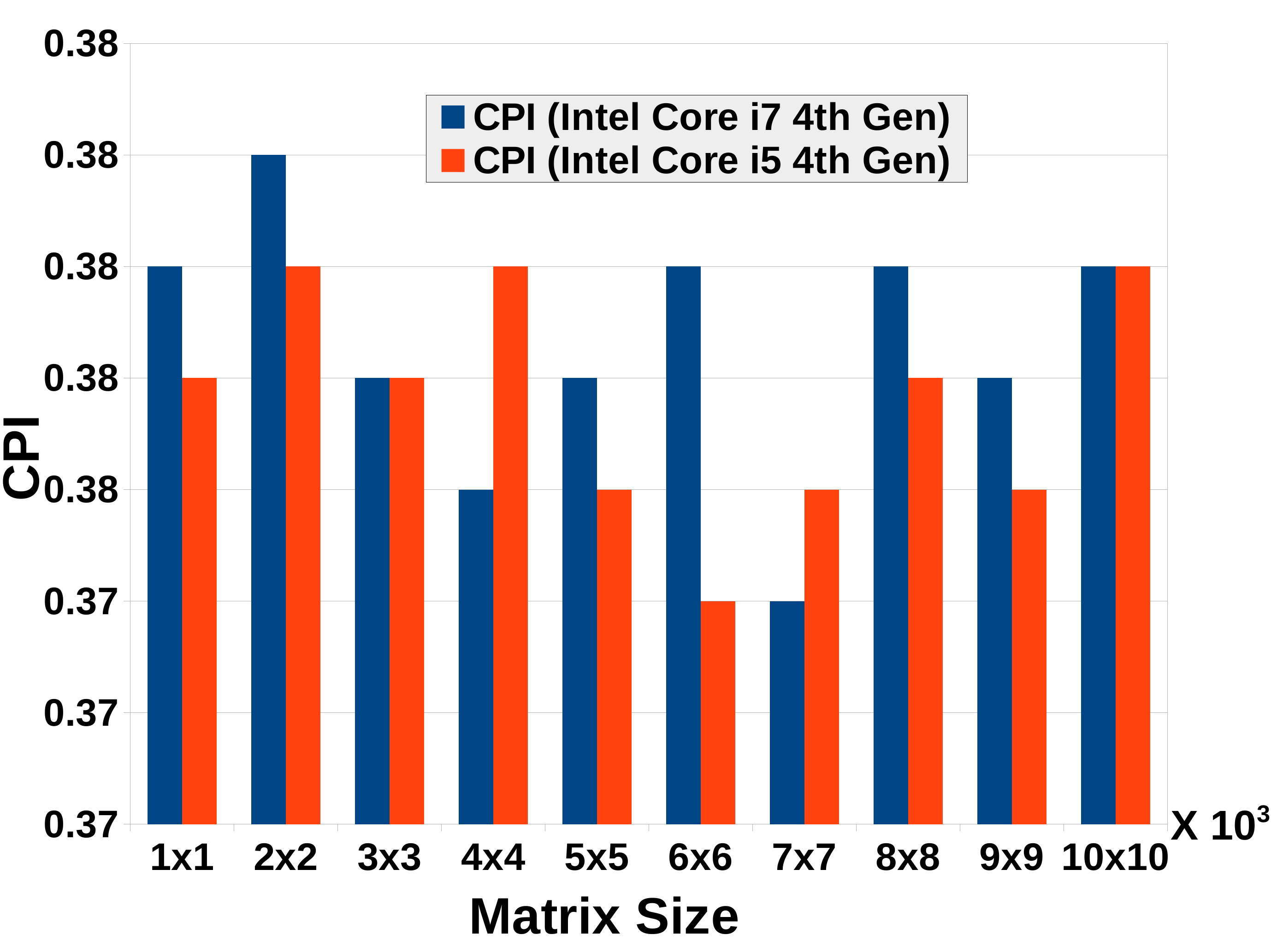}}
\subfigure[Gflops in DGEMM on Intel Haswell Micro-architecture with $icc$ and $-mavx$ compiler switches\label{fig:cpi_gmm_5}]{\includegraphics[scale = 0.18]{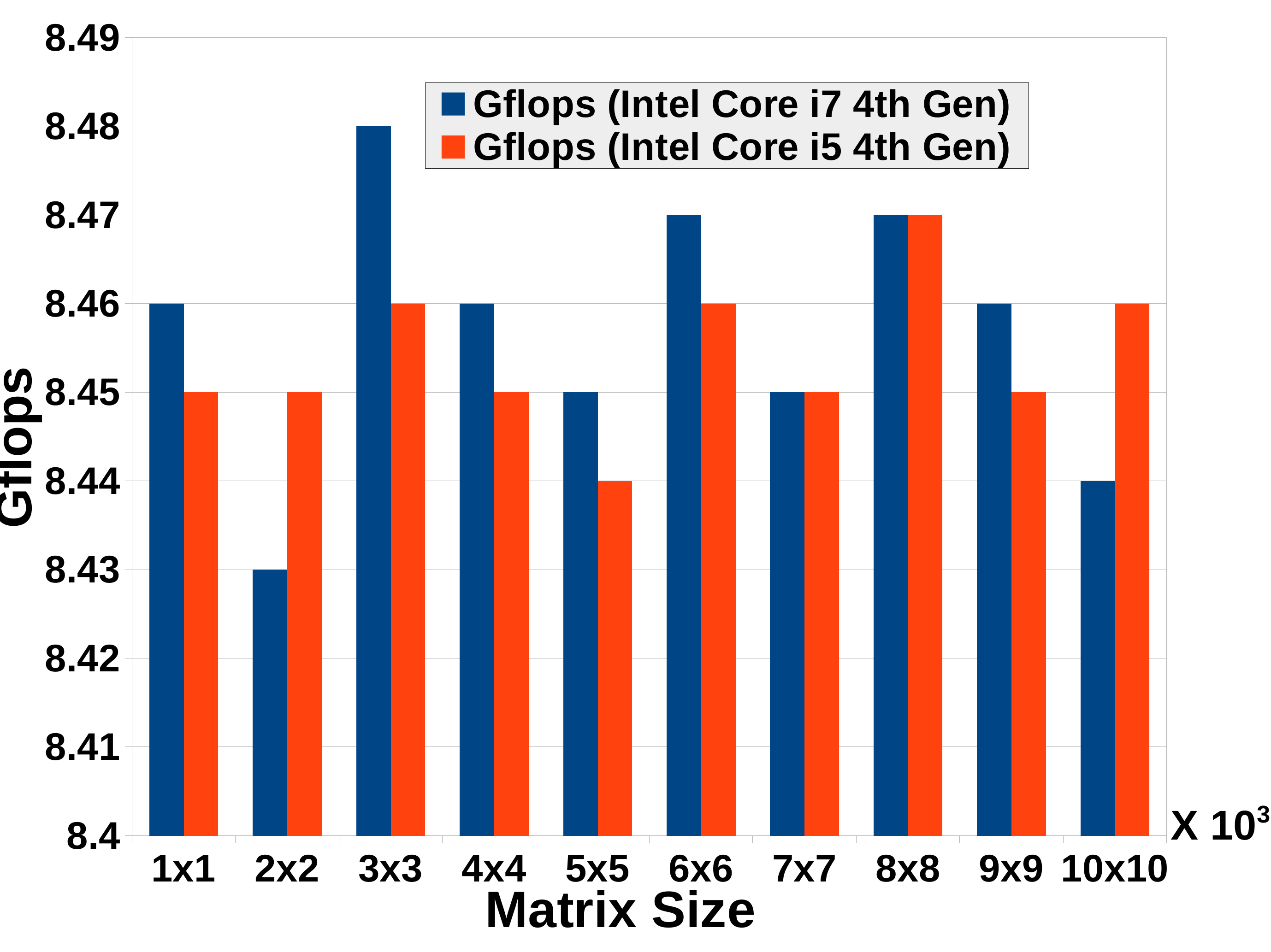}}
\subfigure[Gflops in MAGMA\_DGEMM and MAGMA\_DGEMV on Nvidia Tesla C2050\label{fig:cpi_gmm_6}]{\includegraphics[scale = 0.18]{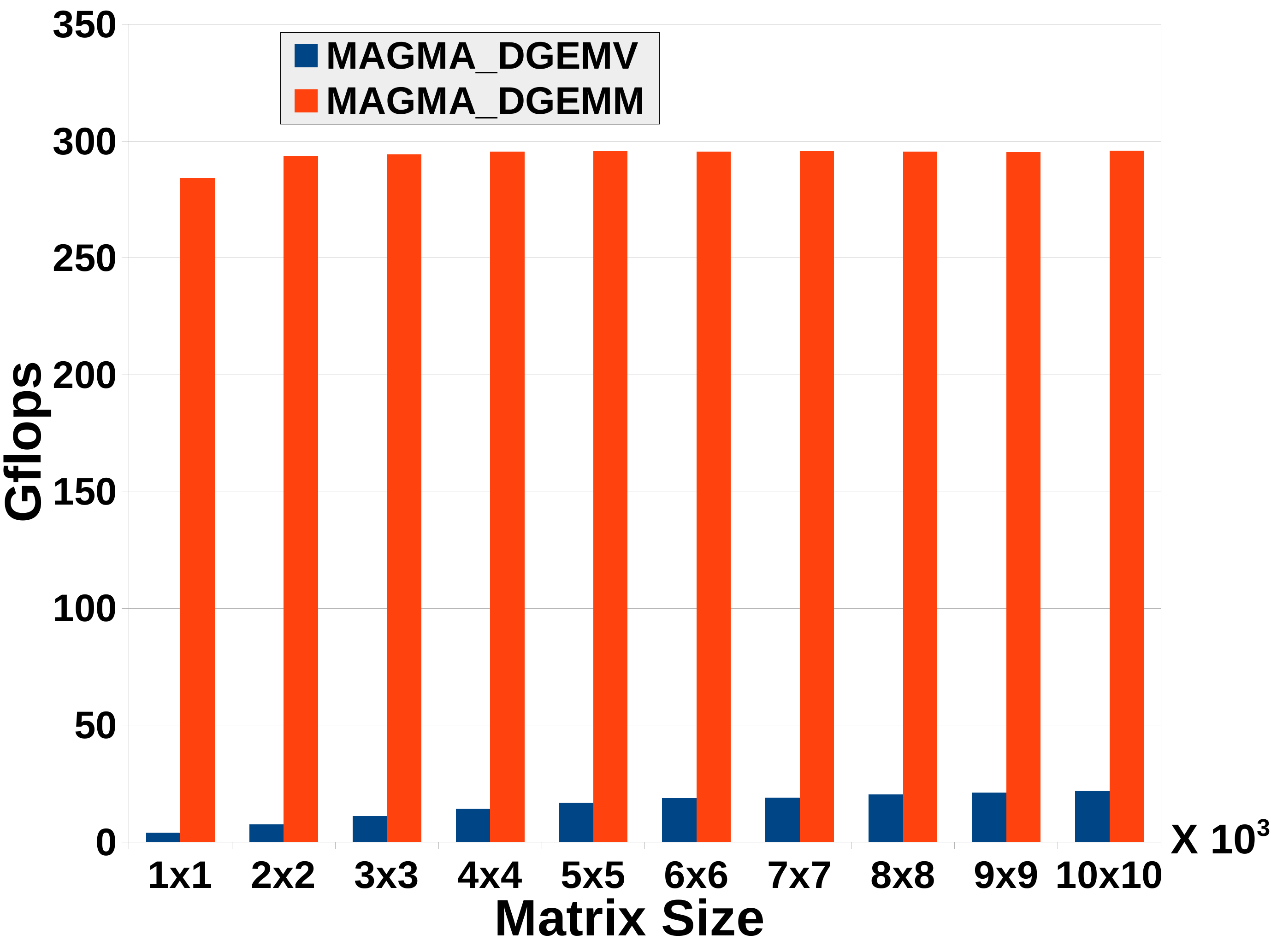}}
\subfigure[Gflops in DGQMV, DGEMM, MAGMA\_DGEMM and MAGMA\_DGEMV \label{fig:cpi_gmm_7}]{\includegraphics[scale = 0.18]{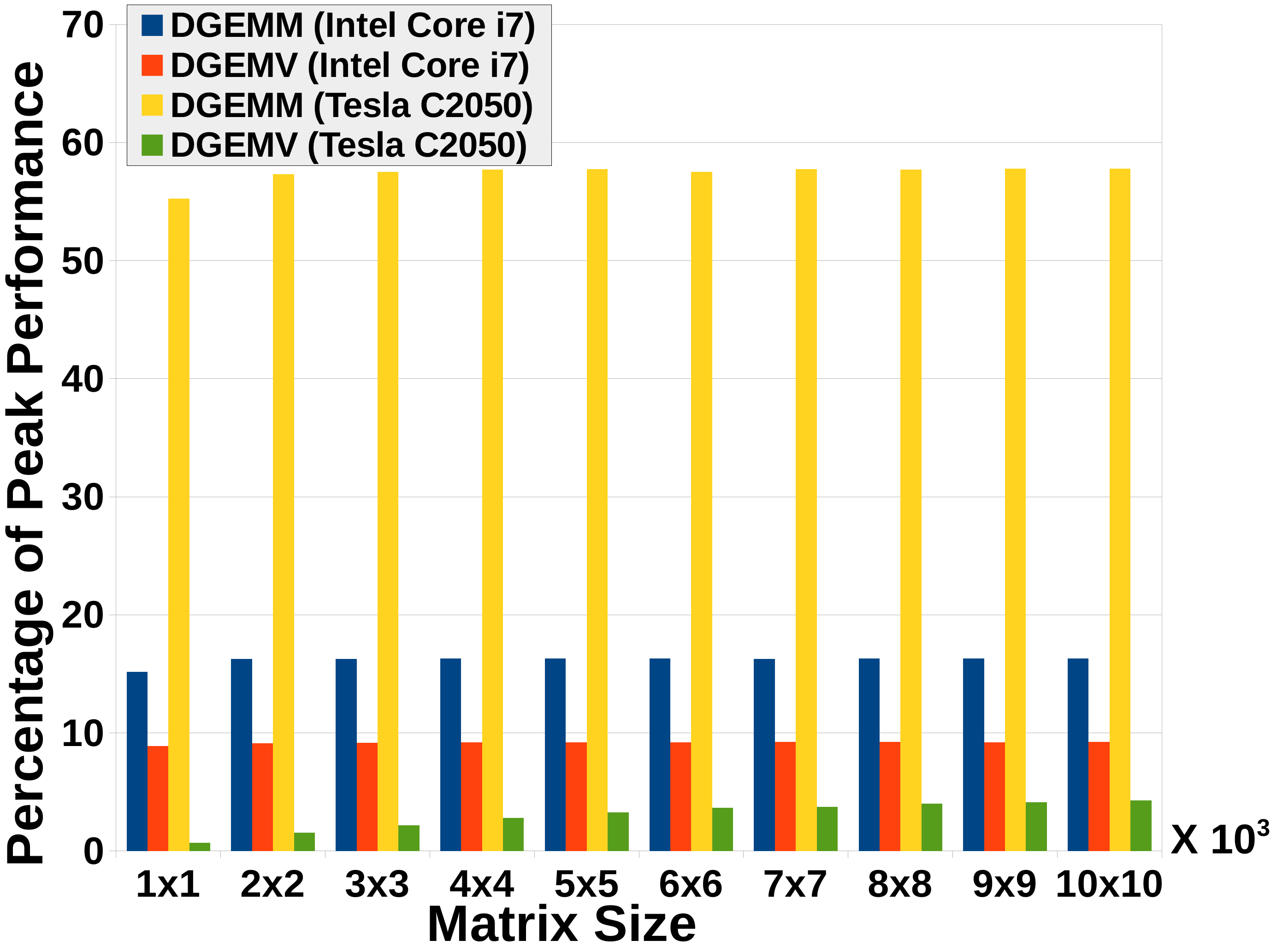}}
\subfigure[Gflops/watt in DGEMV, DGEMM, MAGMA\_DGEMM and MAGMA\_DGEMV on Nvidia Tesla C2050\label{fig:cpi_gmm_8}]{\includegraphics[scale = 0.18]{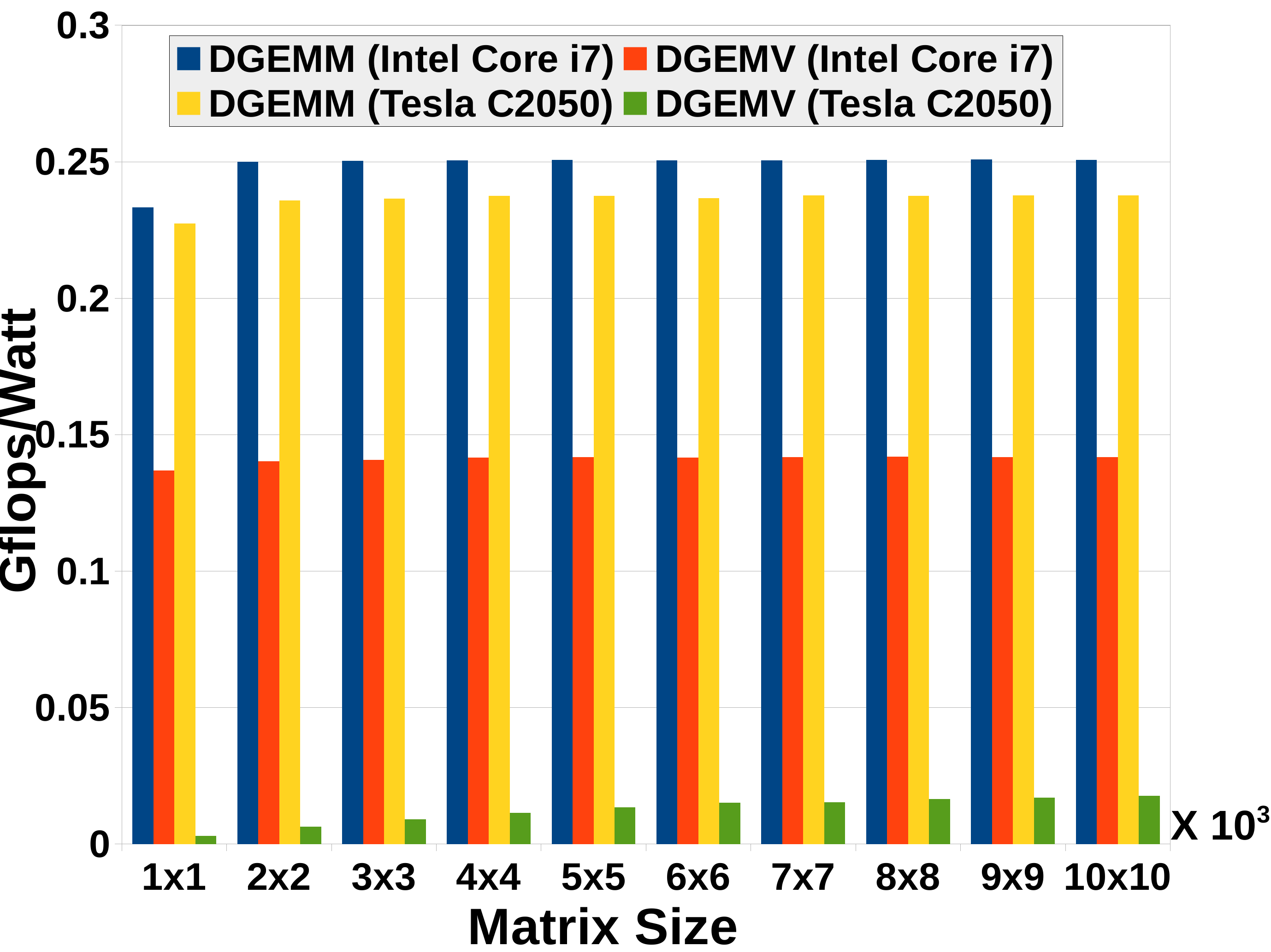}}
\caption{Performance of DGEMV and DGEMM on Different Micro-architectures}
\label{fig:ogr}
\end{figure*}

%
%
%
%
%
%

Figure \ref{fig:cpi_gmm_0} depicts CPI of DGEMM when executed on Intel Haswell and AMD Bulldozer micro-architectures. For experimental results shown in figure \ref{fig:cpi_gmm_0}, we have used BLAS and CBLAS\footnote{CBLAS consists of C wrappers written around BLAS where BLAS is written in Fortran} available in The Netlib and hence we have compiled BLAS and CBLAS using publicly available $gfortran$ and $gcc$. It can be observed in figure \ref{fig:cpi_gmm_0} that the CPI in the DGEMM saturates at around $0.85$ for Intel's Haswell and AMD's Bulldozer. For the matrices that fit in the $L1$ cache achieve CPI that is lower than that for the matrices that do not fit in the $L1$ cache. This is due to $L1$ cache misses observed for larger matrices. While for the matrices that do not fit in the cache memory, attained CPI is slightly higher than the smaller matrices\footnote{Just to re-emphasize: In case of CPI, lower the better}. For Intel Haswell and AMD Bulldozer, the lower bound of the CPI is $0.0625$. It can be observed that with DGEMM, which is highly optimized routine of BLAS, CPI achieved is nowhere close to the lower bound of CPI of the architecture. Similar trend is observed when we consider Gflops as a performance metric as shown in figure \ref{fig:cpi_gmm_1}.

It can be observed in the figure \ref{fig:cpi_gmm_1} that, for the matrices that fit in the cache memory, the Gflops attained is higher. For the larger matrices that do not fit in the cache memory, Gflops decreases due to cache misses. While these architectures have peak performance of $48$ Gflops, attained performance is 10-11\% of the peak performance. 

One way to improve performance is to use the vendor specific compilers, since vendor specific compilers perform architecture aware optimizations in the programs. In order to further push the performance of DGEMM on Intel Haswell micro-architecture we use Intel C Compiler (icc) for compiling DGEMM routine in BLAS. Performance improvement in CPI and Gflops is shown in figures \ref{fig:cpi_gmm_2} and \ref{fig:cpi_gmm_3} respectively.

It can be observed in the figures \ref{fig:cpi_gmm_2} and \ref{fig:cpi_gmm_3} that the performance improvement in DGEMM is still far from the lower bound of the CPI and peak Gflops. 

In the next set of experiments, we add $-mavx$ compiler switch while compiling with $icc$. Performance improvement due to these switches is shown in the figure \ref{fig:cpi_gmm_4} and figure \ref{fig:cpi_gmm_5}.

It can be observed from figures \ref{fig:cpi_gmm_4} and \ref{fig:cpi_gmm_5} that compiler switch $-mavx$ improves performance and finally we are able to achieve 15-17\% of the peak IPC (or CPI) and peak Gflops for DGEMM. Percentage of peak performance achieved is 4-5\% for DGEMV and 55-57\% for DGEMM in Tesla C2050 as depicted in figure \ref{fig:cpi_gmm_6} while percentage of peak performance achieved in Intel and Nvidia machines for DGEMV and DGEMM is ranging from 5\% to 57\% as shown in the figure \ref{fig:cpi_gmm_7}. Considering Gfops/watt as a performance parameter, DGEMV and DGEMM in the BLAS achieve performance of 0.14 Gflops/watt and 0.25 Gflops/watt respectively while MAGMA\_DGEMV and MAGMA\_DGEMM achieve performance of 0.03 Gflops/watt to 0.225 Gflops/watt respectively as shown in figure \ref{fig:cpi_gmm_8}. One more observation we make from the figure \ref{fig:cpi_gmm_4} and VTune\texttrademark$\;\;$that use of $-mavx$ compiler switch along with $icc$ reduces number of instructions by half. This is because of use of FMA instructions in the generated assembly code. Reduction in the number of instructions leads to increase in CPI measured by VTune\texttrademark. Though, there is an increase in CPI, performance in terms of Gflops is observed to be improved as shown in the figure \ref{fig:cpi_gmm_5}. Hence, CPI measured by VTune\texttrademark  can not be considered as a correct measure for performance. We define terms Cycles-per-Flops (CPF) to be used instead CPI and Flops-per-Cycle (FPC) to be used instead IPC as follows:

\begin{align}
	CPF = \frac{Total\;Number\;of\;Clock\;Ticks}{Total\;Number\;of\;Floating\;Point\;Operations}
\end{align}

We define FPC as follows:
\begin{align}
	FPC= \frac{1}{CPF}
\end{align} 

CPF and FPC help us to evaluate performance of the architectures and algorithms more effectively. This is because the granularity of the compute resources considered in CPF and FPC is at the level of floating point operation and not at the level of Fused Multiply-add (FMA).

Across the experiments, we can observe that, significant efforts are needed to improve the performance of DGEMV and DGEMM on contemporary architectures and yet the attained performance is not satisfactory. We address this challenge of extracting performance from DLA computations through algorithm-architecture co-design in the subsequent sections of this paper. 

\section{Analysis of BLAS and CFU}\label{sec:graph}
In this section, we present graph based analysis of several Level-1, Level-2, and Level-3 BLAS. We choose several representative routines in all three levels of BLAS\footnote{Due to availability of double precision floating point unit, we consider only routines that are with prefix "d". For example first "d" in $ddot$ represents double precision}.

\subsection{Vector Operations (Level-1 BLAS)}
\begin{figure*}[]
	\centering
	\includegraphics[scale=0.18]{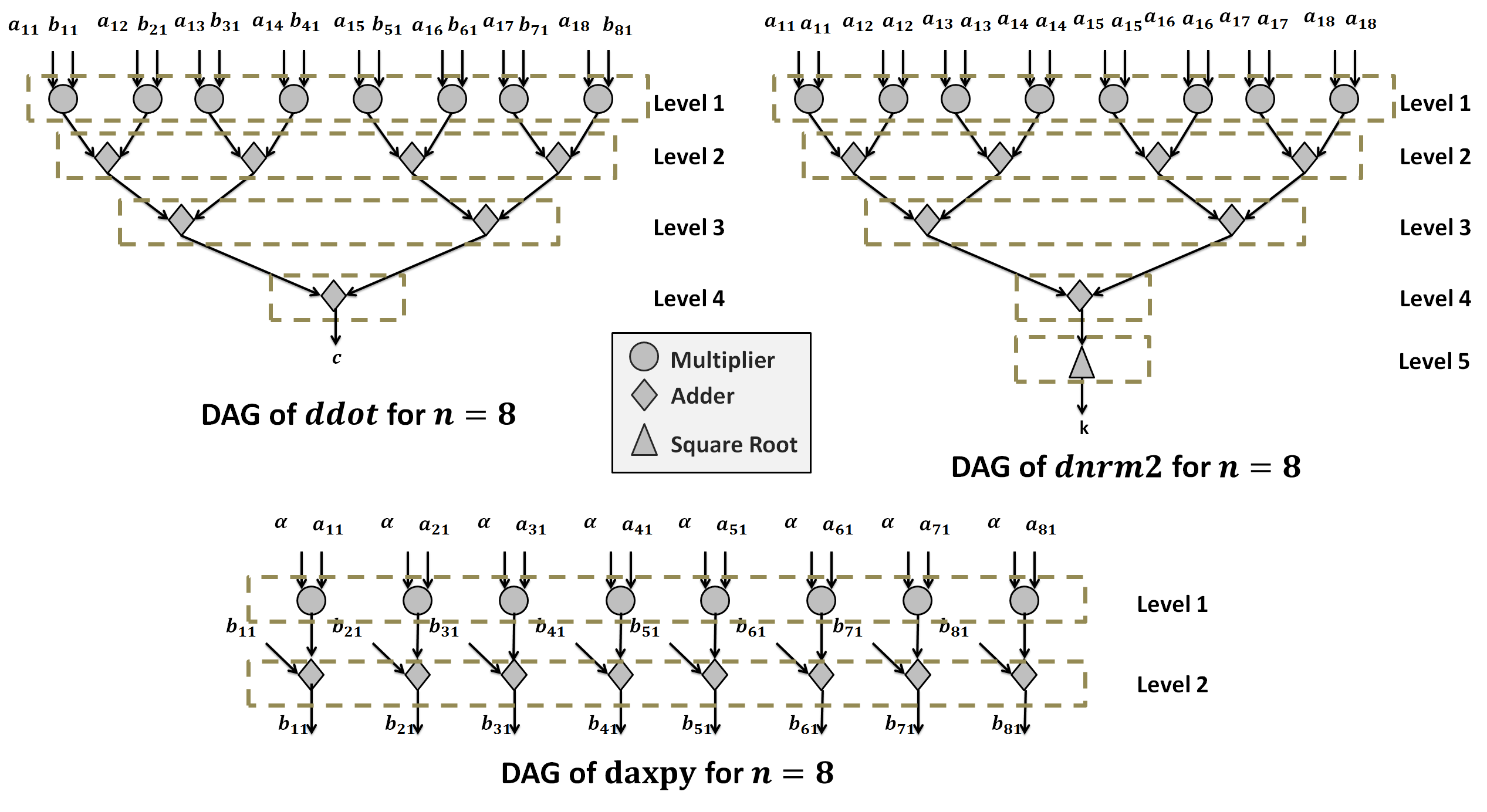}
	\caption{DAGs of $ddot$, $dnrm2$, and $daxpy$ for $n=8$}
	\label{fig:ddot}
\end{figure*}

Level-1 BLAS typically has $O(n)$ operations for a vector size of $n$ and data movement required is also $O(n)$. We analyze $ddot$, $dnrm2$, and $daxpy$ operations here. 
Figure \ref{fig:ddot} represents inner product of two vectors given by equation \ref{eqn:ddot}.

\begin{align}\label{eqn:ddot}
	c = x^Ty
\end{align}
where $x = \begin{bmatrix} a_{11} & a_{12} & a_{13} & ....& a_{1n} \end{bmatrix} $, and $y = \begin{bmatrix} b_{11} \\ b_{21} \\ b_{31} \\ . \\ .\\ . \\ . b_{n1} \end{bmatrix} $. DAG for $ddot$ is shown in figure \ref{fig:ddot} for $n=8$. 

The routine $ddot$ has application in matrix-vector and matrix-matrix multiplication.

$dnrm2$ is given by equation \ref{eqn:drnm2} and DAG for $drnm2$ is shown in figure \ref{fig:ddot}.

\begin{align}\label{eqn:drnm2}
	k = \sqrt{x^Tx}  = \sqrt{a_{11}^2+a_{12}^2+...+a_{1n}^2}
\end{align}


$daxpy$ is given by equation \ref{eqn:daxpy} and DAG for $daxpy$ is given figure \ref{fig:ddot}

\begin{align}\label{eqn:daxpy}
	y = \alpha x + y
\end{align}


It can be observed from the DAGs of $ddot$ and $dnrm2$ that the DAGs of these two routines are similar except presence of square root in $dnrm2$ routine. $dnrm2$ can be realized with same multiplier and adder resources as $ddot$. It can also be observed in the figure \ref{fig:ddot} that the first level in the DAGs is multiplication and all these multiplications can potentially be executed in parallel. The next levels of DAGs of $ddot$ and $dnrm2$ are additions. Additions in each level of DAGs can be performed simultaneously if the inputs are available.  

\subsection{Matrix-vector Operations (Level-2 BLAS)}

\begin{figure}[]
	\centering
	\includegraphics[scale=0.22]{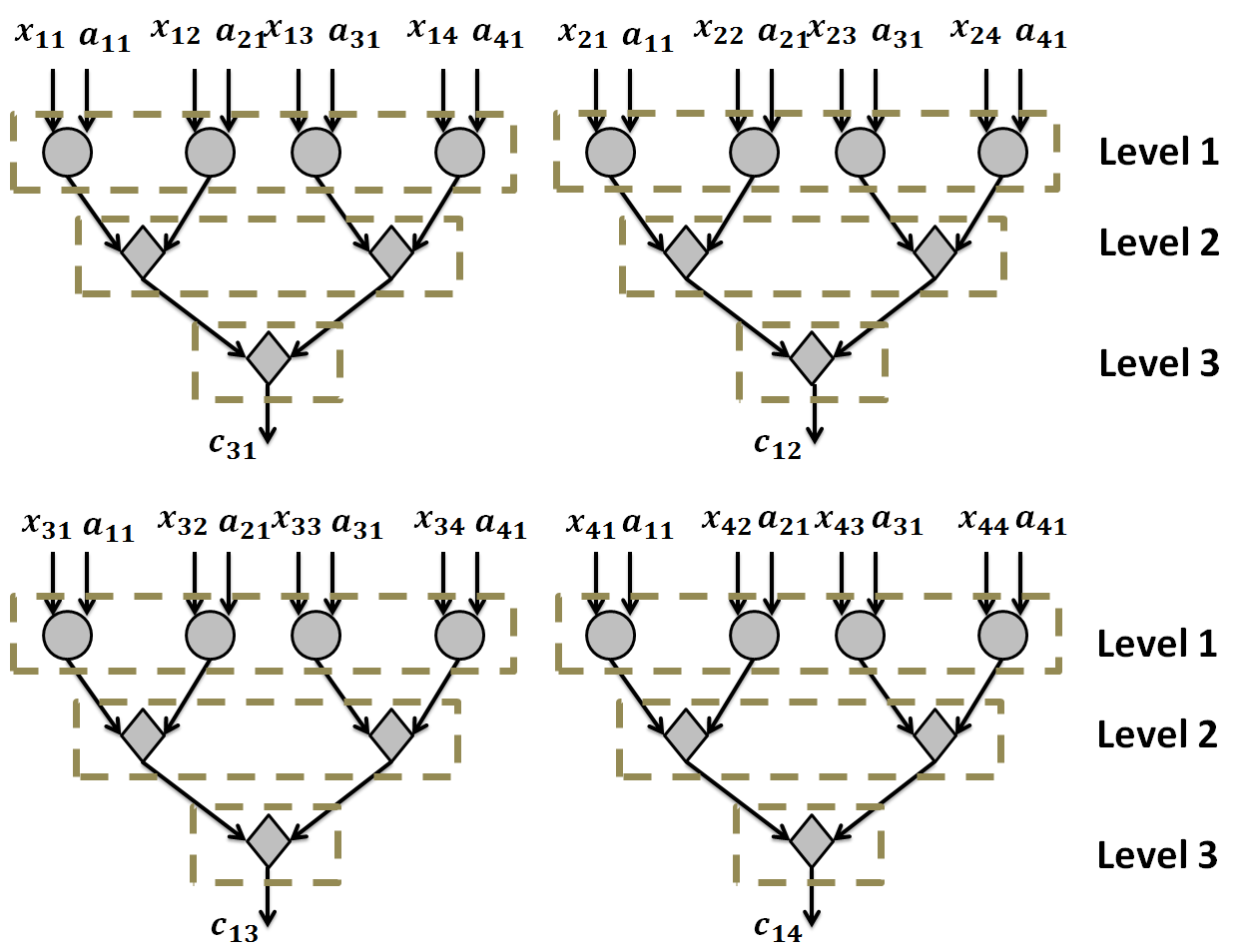}
	\caption{DAG of DGEMV for $n=4$}
	\label{fig:dgemv}
\end{figure}

Matrix-vector operations are typically $O(n^2)$ for input matrix of size $n\times n$ and vector of size $n$ and data movement required is also $O(n^2)$.  In our analysis we consider double precision matrix-vector multiplication (DGEMV) routine of BLAS. DGEMV is given in equation \ref{eqn:dgemv}.

\begin{align}\label{eqn:dgemv}
	y = Ax + y
\end{align}
where $A$ is a matrix of size $n\times n$ and $x$ and $y$ are vectors of size $n$. DGEMV routine has $n^2$ multiplications, $n^2-n$ additions and $n$ additions to compute final vector $y$. In our DAG analysis we consider matrix-vector multiplication since matrix-vector multiplication is compute intensive part of the routine and it is essential to exploit parallelism in matrix-vector multiplication to accelerate DGEMV routine. DAGs for matrix-vector multiplication are shown in in the figure \ref{fig:dgemv}.

It can be observed from DAGs of matrix-vector multiplication that all the multiplication in matrix-vector multiplication can be executed in parallel and matrix-vector multiplication can be realized as series of $ddot$ routine calls. 

\subsection{Matrix-matrix Operations (Level-3 BLAS)}

Here we first review some of the matrix multiplication algorithms and present graph based analysis  of these algorithms. We briefly discuss parallelism in different matrix multiplication algorithms. Based on the analysis, we choose matrix multiplication algorithm. In the following subsection of the paper, we present design of a PE that can efficiently exploit ILP in the chosen matrix multiplication algorithm.

\subsubsection{Matrix Multiplication Algorithms}
Over the years there have been several matrix multiplication algorithms proposed in the literature. In this subsection, we review and analyze Strassen's Matrix Multiplication (SMM), Winograd's Matrix Multiplication (WMM), and General Matrix Multiplication (GEMM). We consider $  A=\begin{bmatrix} A11 & A12 \\ A21 & A22 \end{bmatrix}$, and $  B=\begin{bmatrix} B11 & B12 \\ B21 & B22 \end{bmatrix}$ as input matrices and $  C=\begin{bmatrix} C11 & C12 \\ C21 & C22 \end{bmatrix}$ as output matrix where $A,B,C$ are equal sized block matrices and $A,B,C$ $\in$ $\R^{2n}\times \R^{2n}$. 
	\subsubsection{Strassen's Matrix Multiplication}

SMM algorithm is described in table \ref{tab:smm} for $2\times 2$ block matrix. 

\begin{table*}[]
\centering
\caption{Computations in the different Levels of DAGs of SMM at first step of Recursion in $2\times 2$ Block Matrix Multiplication \label{tab:exp1}}{
\begin{tabular}{|c|p{2.5cm}|p{2.2cm}|p{2.2cm}|}\hline
Level 1 & Level 2  & Level 3 & Level 4\\ \hline\hline
$T1 = A11 + A22$ & $M1 = T1T2$ &  $K1 = M1+M4$ & $C11 = K1 - K2$ \\ \hline
$T2 = B11 + B22$ & $M2 = T2B11$ & $K2 = M3 - M7$ &  $C22 = K3 + K4$ \\ \hline
$T3 = B12 - B22$ & $ M3 = A11T3$ & $K3 = M1 - M2$ &  \\ \hline
$T4 = B21 - B11$ & $M4 = A22T4$ & $K4 = M3+M6$ &  \\ \hline
$T5 = A11 + A12$ & $M5 = T5B22$ & $C12 = M3+M5$ &  \\ \hline
$T6 = A21 - A11$ & $M6 = T6T7$ & $C21 = M2+M4$ &  \\ \hline
$T7 = B11 + B12$ & $M7 = T8T9$ &  &  \\ \hline
$T8 = A12 - A22$ &              &  &  \\ \hline
$T9 = B21 + B22$ &              &  &  \\ \hline
\end{tabular}}
\label{tab:smm}
\end{table*}

\begin{figure*}[]
	\centering
	\includegraphics[scale = 0.16]{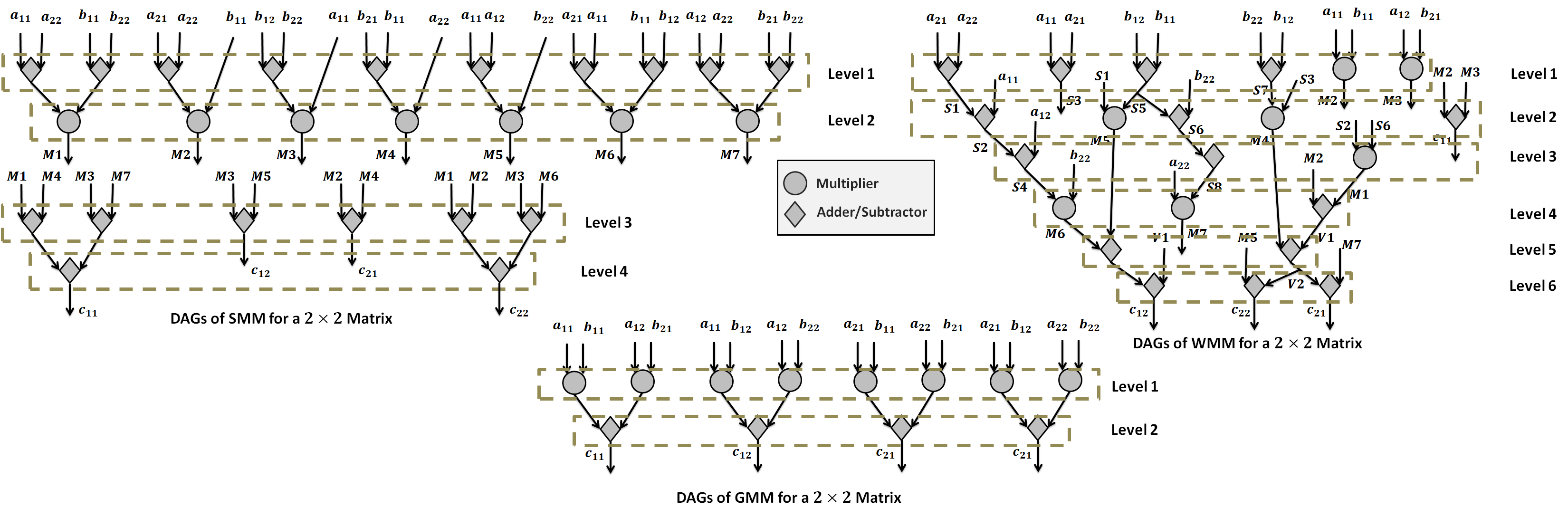}
	\caption{SMM, WMM, and GEMM for $2\times 2$ Block Matrix}
	\label{fig:smm1}
\end{figure*}

Typically, SMM has two steps, 1) decompose step, and 2) merge step. In decompose step, matrix is divided in block matrices and $M1$ to $M7$ are computed. In merge step, $C11$ to $C22$ are computed. Directed Acyclic Graphs (DAGs) for SMM are shown in figure \ref{fig:smm1} for $n=1$. It can be observed in the DAGs of SMM that the computation of $C11$ to $C22$ depends on computations of $M1$ to $M7$.  

This dependencies in the DAGs of SMM results in higher execution time of SMM. It can also be observed from DAGs of SMM in figure \ref{fig:smm1} that M1 to M7 can potentially be executed in parallel and C11 to C22 can also be executed in parallel. One of the way this parallelism can be exploited is through processor pipeline and pipelined arithmetic units. Asymptotic complexity of SMM is $O(n^{2.81})$.

	\subsubsection{Winograd's Matrix Multiplication}
		WMM algorithm operates on the same principle as SMM as shown below:

\begin{table*}[]
\centering
\caption{Computations in the different Levels of DAGs of WMM at first step of Recursion in $2\times 2$ Block Matrix Multiplication \label{tab:exp1}}{
\begin{tabular}{|c|p{2.5cm}|p{2.5cm}|p{2.5cm}|p{2.5cm}|p{2.5cm}|p{2.5cm}|}\hline
Level 1 & Level 2  & Level 3 & Level 4 & Level 5 & Level 6\\ \hline\hline
$S1 = A21 + A22$  & $S2 = S1 - A11$  & $S4 = A12 - S2$ & $M6 = S4B22$ & $V2 = V1 + M4$ & $C12 = V1 + K1$\\ \hline
$S3 = A11 - A21$ & $S6 = B22 - S5$ & $S8 = S6 - B21$ & $M7 = A22S8$ & $K1 = M5 + M6$ & $C21 = V2 - M7$ \\ \hline
$S5 = B12 - B11$ & $M4 = S3S7$ & $M1 = S2S6$ & $V1= M1 + M2$ &  & $C22 = V2 + M5$ \\ \hline
$S7 = B22 - B12$ & $M5 = S1S5$ &  & &  &  \\ \hline
$M2 = A11B11$ & $C11 = M2 + M3$ &  & &  &  \\ \hline
$M3 = A12B21$ &  &  & &   & \\ \hline
 \end{tabular}}
\end{table*}



It can be observed from WMM algorithm that it takes $7$ block matrix multiplications and $15$ matrix additions unlike SMM where the number of block matrix multiplication is same but the number of matrix additions are $18$. DAGs for WMM are shown in figure \ref{fig:smm1} for $n=1$.

WMM has same asymptotic complexity as SMM. In practical scenarios, execution time of WMM is observed to be slightly less than SMM due to fewer additions.

	\subsubsection{General Matrix Multiplication}
GEMM for multiplication of $A$ and $B$ can be described by following expressions:

\begin{align}
	C11 &= A11B11 + A12B21 \nonumber \\
	C12 &= A11B12 + A12B22 \nonumber \\
	C21 &= A21B11 + A22B21 \nonumber \\
	C22 &= A21B12 + A22B22 \nonumber \\
	\nonumber
\end{align}


DAGs for a $2\times 2$ matrix is shown in figure \ref{fig:smm1}. It can be observed from the DAGs of GEMM that it takes $8$ multiplications and $4$ additions. Asymptotic complexity of GEMM is $O(n^3)$.

SMM and WMM have lower asymptotic complexities compared to GEMM. A major disadvantage of SMM and WMM is that they are more suitable for square matrices where size is a power of two. For the matrix sizes where this condition is not met, a complex matrix partitioning scheme is required. Hence, we adopt GEMM over SMM and WMM due to following reasons:

\begin{itemize}
	\item A complex partitioning scheme required for the matrices in SMM and WMM results in a intricate scheduling scheme for the blocks of input matrices. A way to alleviate these complications is to $zero$ pad the matrices. This $zero$ padding results in few more computations, mostly $O(n^2)$. The $zero$ padding does not reduce the complexity of the implementation since naive $zero$ padding scheme is not efficient 
	\item GEMM has higher pedagogical importance than SMM and WMM. GEMM is highly preferred to evaluate the emerging architectures over SMM and WMM due to its simple structure and ease of implementation
\end{itemize}

\subsubsection{Anatomy of General Matrix Multiplication}
To discuss available parallelism in GEMM, we take a matrix multiplication of size $4\times 4$ as an example. 

\begin{figure}[]
	\centering
	\includegraphics[scale = 0.14]{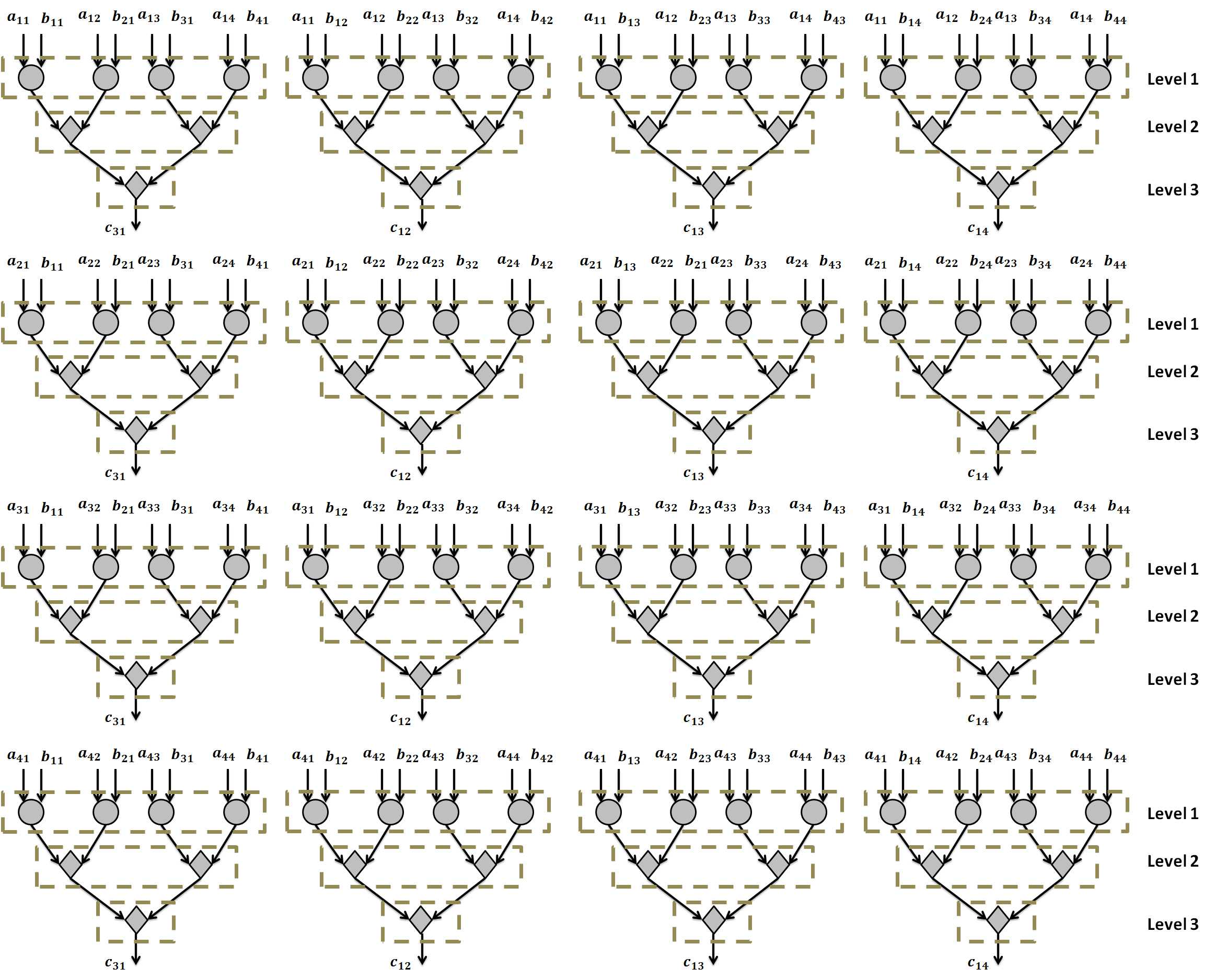}
	\caption{DAGs in GEMM for $4\times 4$ Matrix}
	\label{fig:gmm_dag}
\end{figure}

DAGs for $m=n=4$ for algorithm \ref{algo:gmm1} are shown in figure \ref{fig:gmm_dag} for computation of elements $c11$ to $c44$. It can be observed in the figure \ref{fig:gmm_dag} that all the multiplications in the block of the matrix can be computed in parallel. The dependencies are due to accumulation of the multiplied elements of the input matrices. 

Potentially, in multiplication of matrix of size $n\times n$, all the $n^3$ multiplications can be computed in parallel. In case of $4\times 4$ matrix, all $16$ elements can be computed in parallel as shown in figure \ref{fig:gmm_dag}. The accumulation process while computing the elements of the resultant matrix enforce the dependencies resulting in pipeline stalls. These pipeline stalls can be avoided by computing multiple elements in parallel-pipeline manner.

\begin{algorithm}
\caption{Block General Matrix Multiplication}
\label{algo:gmm_unrol1}
\begin{algorithmic}[1]
\State Allocate memories for input and output matrices
\For{$i=1$ to $m/4$}
  \For{$j=1$ to $n/4$}
    \For{$k=1$ to $n/4$}
      \State C = BLOCK4ADD(BLOCK4MUL(A,B),C)
     \EndFor
  \EndFor
\EndFor
\end{algorithmic}
\end{algorithm}


Algorithm \ref{algo:gmm_unrol1} depicts DGEMM with $4\times 4$ block matrix multiplication (assuming that the matrix dimensions are multiple of $4$). In algorithm \ref{algo:gmm_unrol1}, BLOCK4MUL is multiplication of matrices of size $4\times 4$, and BLOCK4ADD is addition of matrices of size $4\times 4$.
A pitfall in the unrolling scheme is the exigency of locally available registers. Typically, for $n\times n$ matrix, if fully unrolled, requires $3n^2$ registers. Hence, for a large matrix, it can not be unrolled due to lack of locally available registers, but a small block of the matrix can be unrolled to exploit the fine grained parallelism in the block through processor pipeline and pipelined arithmetic units. In our experiments with PE explained in section \ref{sec:pe_design}, we have adopted a conservative approach where we have assumed space for $n^2$ intermediate results in the loacal registers and hence we have considered a $4\times 4$ block matirx with 64 registers of 64-bit wide.    In parallel realization of GEMM, different blocks of $4\times 4$ can be computed in parallel on different PEs. While realizing GEMM on a single PE, we try to exploit parallelism that is available in a block of $4\times 4$ and in parallel realization on REDEFINE, we try to exploit parallelism across the blocks. 

In the next section we present a PE design to skillfully exploit the parallelism that exist in the block of $4\times 4$ matrix.

\subsection{Processing Element Design}\label{sec:pe_design}

\begin{figure}[]
	\centering
	\includegraphics[scale = 0.35]{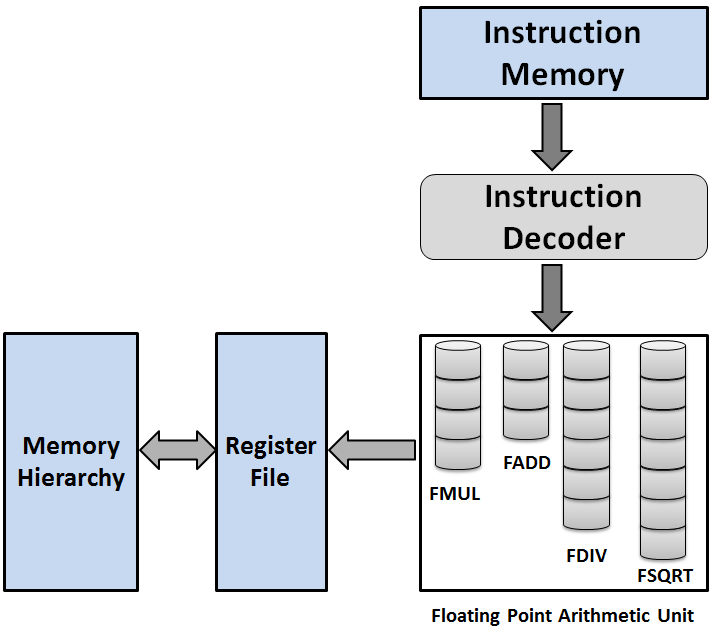}
	\caption{PE Architecture}
	\label{fig:fps}
\end{figure}

For initial design of PE, we consider classical sequential architecture model. As a first design, we take a fully pipelined double precision floating point adder, and multiplier arithmetic units as compute resources. Architecture of PE is shown in figure \ref{fig:fps}. 

As shown in the figure \ref{fig:fps}, the initial design of PE consists of an Instruction Memory, a Decoder to decode the instructions, a Register File with $64$ registers, and pipelined double precision Floating Point Unit (FPU) \cite{fpu2}\cite{fpu3}. The FPU consists of a multiplier, an adder, a square root, and a divider.  For computing matrix multiplication with large matrices, we choose $4\times 4$ as a block matrix. For the matrices that are not multiple of $4$, we partition them in the blocks of $4\times 4$ as many times as possible and for the residual matrix, we perform unblocked multiplication. Operation of the PE can be described in the following steps:

\begin{itemize}
	\item {\bf Step 1:} Bring the input matrices to the Register File that by sending Load request to the upper level of the memory
	\item {\bf Step 2:} Perform matrix multiplication
	\item {\bf Step 3:} Store back the resultant matrix to the upper level of memory
\end{itemize}

	\subsection{Simulation Environment and Initial Results}

For simulations we connect PE shown in the figure \ref{fig:fps} to external/global memory. Initially we use 64 ($64-bit$ wide) registers, 16KB of instruction memory for our experiments. We model global/external memory delay by placing pipelined delay of $20$ stages. 

\subsubsection{Initial Results}
For our experiments, we choose matrix sizes $20\times 20$, $40\times 40$, $60\times 60$, $80\times 80$, and $100\times 100$ as a representative matrix sizes for our experiments. 

\begin{table}[]
\centering
\caption{Latencies, CPF and Efficiency \label{tab:exp1}}{
\begin{tabular}{|p{1.4cm}|p{2.8cm}|p{0.45cm}|p{0.9cm}|p{1.2cm}|}\hline
Matrix Size & Experimental Latencies  & CPF & Gflops/W  \\ \hline\hline
$20\times 20$ & 39000 &  1.625 & 16.66  \\ \hline
$40\times 40$ & 310075 & 1.614 & 16.87  \\ \hline
$60\times 60$ & 1040754 & 1.606 & 17.15  \\ \hline
$80\times 80$ & 2457600 & 1.6 & 17.25 \\ \hline
$100\times 100$ & 4770000 & 1.59 & 17.38  \\ \hline
\end{tabular}}
\end{table}

It can be observed in the table \ref{tab:exp1} that as we increase the matrix size, the CPF decreases and saturates around $1.6$ while performance in terms of Gflops/watt is observed to be at 17.3 Gflops/watt at 0.2 GHz. In other words, as we increase the matrix size, the FPC saturates at 62.5\% of peak floating point operations per cycle. Here, since, we can potentially compute one multiplication and one addition in parallel, the peak FPC = 2.

In this section, we reviewed some of the matrix multiplication techniques. We discussed asymptotic complexity and graph based analysis of three different matrix multiplication algorithms. We justified our choice of GEMM over SMM and WMM algorithms. We presented additional details of GEMM with an example of $4\times 4$ matrix and proposed an initial design of a PE that achieves CPF of $1.6$ and performance of 17.3 Gflops/watt. Intuitively, performance of the PE can be improved methodically by enriching the PE with compute and memory resources.

\section{Micro-architectural Enhancements in PE and Parallel Realization on REDEFINE}\label{sec:real}
Based on anatomy of the GEMM and design of the PE presented in section \ref{sec:pe_design}, in this section we dwell on micro-architectural enhancements of the PE. We methodically enhance PE that improves CPF. The PE described in section \ref{sec:pe_design} is considered as a Floating Point Sequencer in this section. 

\subsection{Introducing Local Memory and Load-Store CFU}
\begin{figure}[]
	\centering
	\includegraphics[scale=0.20]{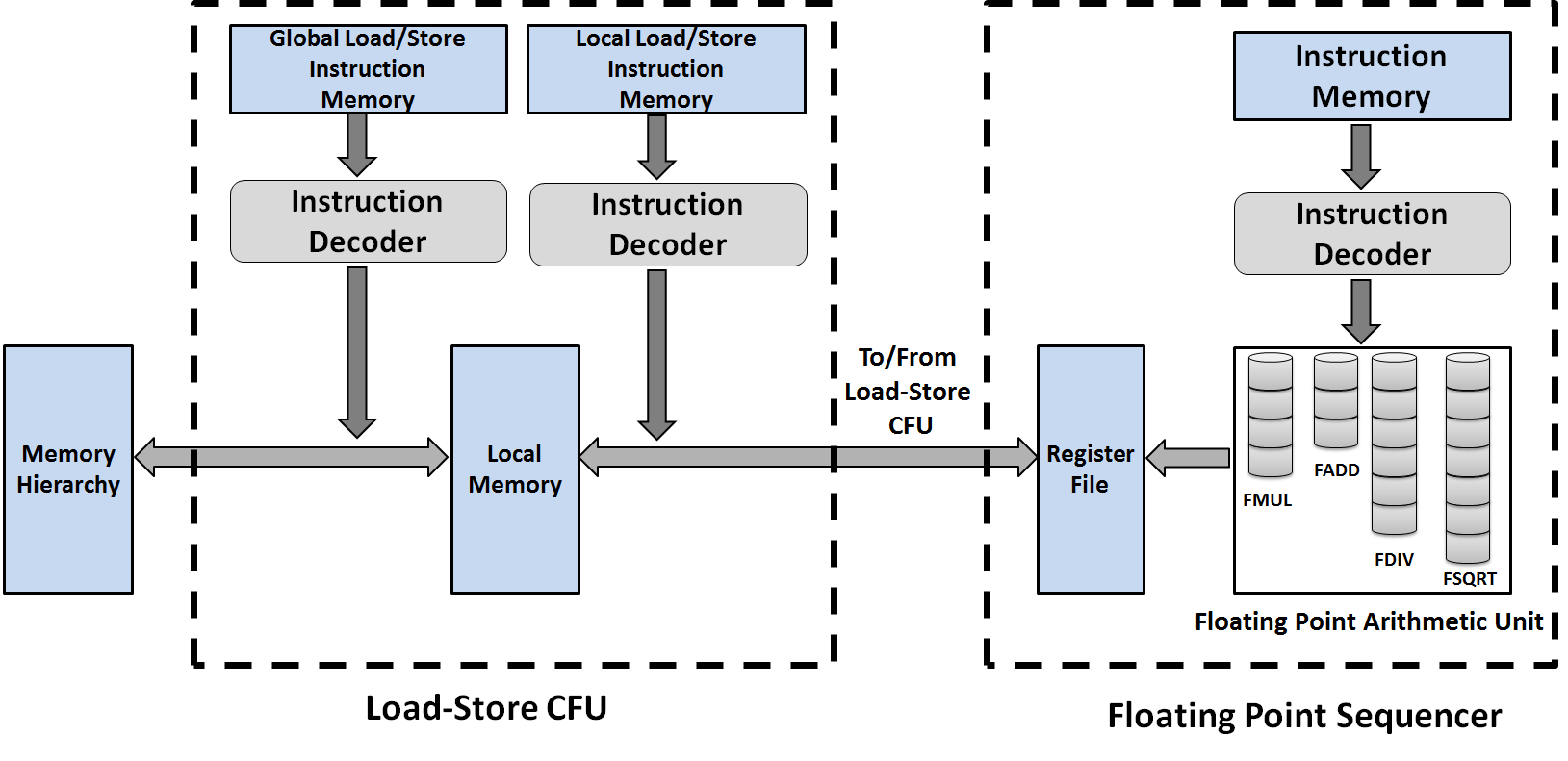}
	\caption{PE with FPS and Load-Store CFU}
	\label{fig:cfu1}
\end{figure}

Major drawbacks of the PE design presented in section \ref{sec:pe_design} are no overlap of computations and communication and lack of exploitation of data locality in GEMM. To address these issues, we introduce a Load-Store CFU that operates simultaneously with FPS (depending on availability of data), and facilitates overlap of computation and communication. We also place a Local Memory (LM) of size $256$kbits in the Load-Store CFU to exploit data locality in GEMM. Enhanced PE design with FPS and Load-Store CFU is shown in figure \ref{fig:cfu1}.  

Introduction of an LM in Load-Store CFU inside PE improves the data locality. This improved data locality creates further opportunities for exploitation of higher ILP by increasing compute resources in FPS. Increased compute resources in the FPS demand for improvisation in the data availability in the Register File for computations. In this section we present methodical architectural enhancements in the PE for reduction in latency\footnote{Latency in terms of clock cycles} in execution of GEMM. These enhancements in the PE ensures lower latency in execution of GEMM leading to overlap between computation and communication up-to 90\% in GEMM and we are also able to achieve up-to 74\% of the peak CPF\footnote{Peak CPF =$ \frac{1}{Number\;of\;Arithmetic\;Units\;that\;can\;function\;concurrently}$}. To highlight the reduction in the latency due to each architectural enhancement, we consider $20\times 20$, $40\times 40$, $60\times 60$, $80\times 80$, and $100\times 100$ matrix sizes as a representative for our experiments. Reduction in the latency due to introduction of Load-Store CFU and LM (refer figure \ref{fig:cfu1}) is shown in the table \ref{tab:mm2}. 

\begin{table*}[]
\centering
\caption{Latencies of $20\times 20$, $40\times 40$, $60\times 60$, $80\times 80$, and $100\times 100$ GEMM (with LM and Load-Store CFU (PE-Architectural Enhancement $1$ (AE1))\label{tab:mm2}}{
\begin{tabular}{ |p{6cm}|c|c|c|c|c| } 
 \hline
 Matrix Size & $20\times 20$ & $40\times 40$ & $60\times 60$ & $80\times 80$ & $100\times 100$\\  \hline \hline
Latency (in clock cycles) without LM & 39000 & 312075 & 1040754  & 2457600 & 4770000
 \\ \hline 
 Latency (in clock cycles) with LM & 23000 & 178471 & 595421 & 1410662 & 2730365
 \\ \hline 
Improvement in Latency in terms of percentage & 41\% & 42.5\% & 42.78\% & 42.6\% & 42.6\% \\ \hline
Gflops/watt & 14.87 & 15.53 & 15.77 & 15.81 & 15.98 \\ \hline
\end{tabular}}
\end{table*}

It can be observed in the table \ref{tab:mm2} that introduction of LM in the Load-Store CFU improves performance by $2$x and as we increase matrix size the performance improves due to improved data locality. 

\subsection{Special Instructions}
In the first enhancement, we try to exploit higher ILP by increasing resources in the FPS that in turn improves performance significantly. This improved performance motivates us to improve data availability in the Register File residing inside FPS. 
We introduce two types of special instruction: 1) DOT instructions that are executed in FPS on a specialized fully pipelined hardware structure, and 2) Block Data Load and Block Data Store instructions that are executed in the Load-Store CFU.
	\subsubsection{DOT Instruction}\label{sec:si}
\begin{figure}[]
	\centering
	\includegraphics[scale=0.14]{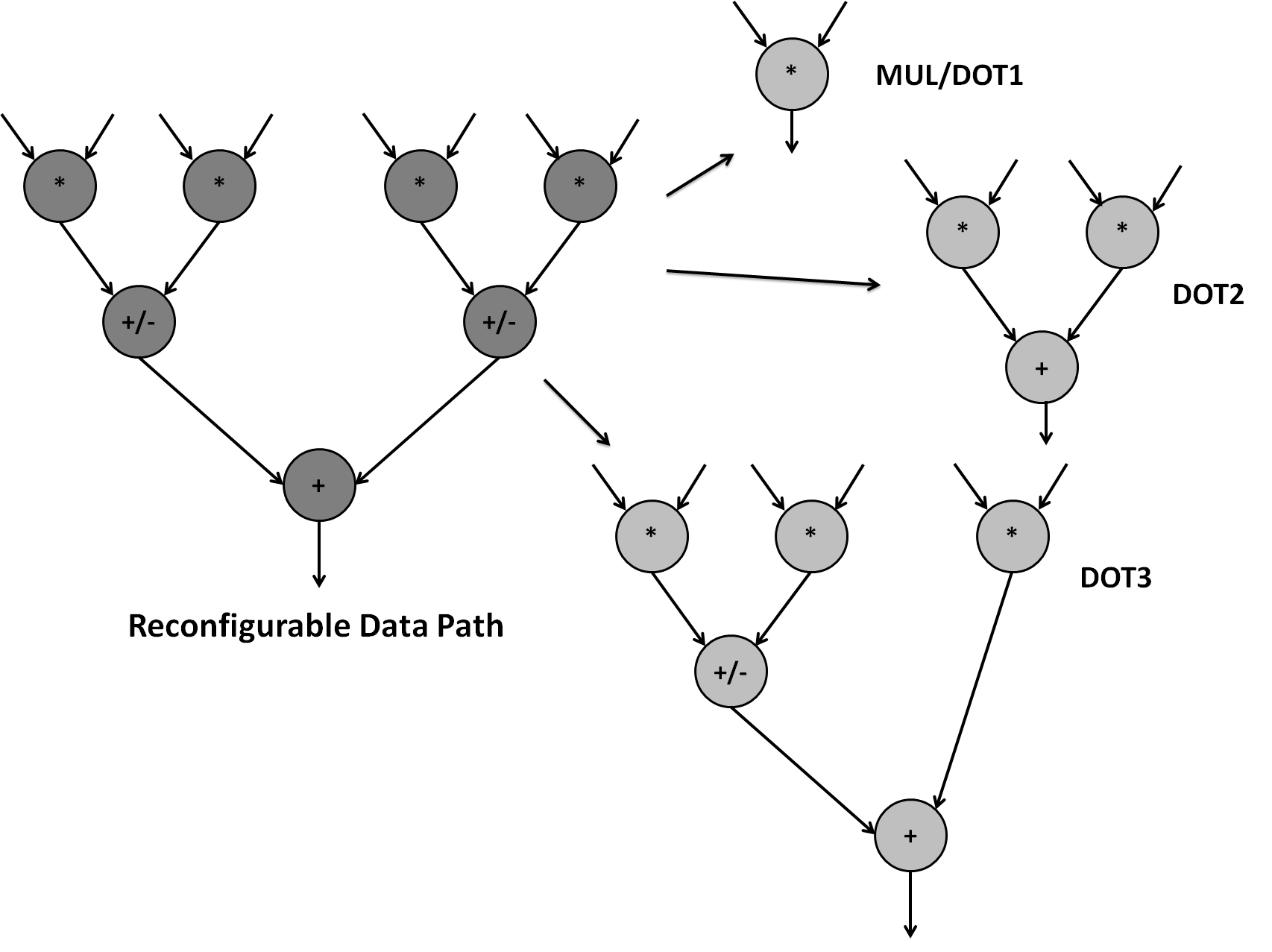}
	\caption{RDP of DOT Instruction and different data-path derived by different configurations of DOT}
	\label{fig:dot}
\end{figure}

	Since we support block size of $4\times 4$, we introduce a hardware that can perform inner product of a $4$-element vector. The hardware structure to compute $4$-element vector inner product is shown in figure \ref{fig:dot}. We further make this hardware structure reconfigurable to support $2$-element and $3$-element vector inner products to support different matrix sizes. We name this unit as a Reconfigurable Data-path (RDP). Through reconfiguration, RDP can be re-casted to perform macro operations encountered in some of the algorithms in BLAS discussed in the section \ref{sec:graph}. The RDP is shown in figure \ref{fig:dot}.

For larger matrices ($>\;4\times 4$) that do not fit in the Register File in FPS for matrix multiplication, we use block size of $4\times 4$. For the matrices that do not have their size as multiple of $4$, we use $2$-element, and $3$-element inner product configurations of RDP. In this exposition, we restrict our experiment to the matrices with size of multiple of $4$ and hence we use $4$-element inner product configuration (also termed as DOT4 configuration) of RDP. 
DOT4 configuration of RDP has a $15$-stage deep pipeline. Assuming no pipeline stalls, we can potentially maintain $15$ DOT4 instructions in a state of execution. This DOT4 instruction leads to exploitations of higher ILP in a block of $4\times 4$ in GEMM. Improvement in latency of GEMM due to DOT4 instruction is shown in table \ref{tab:mm3}.

\begin{table*}[]
\begin{center}
\caption{Latencies of $20\times 20$, $40\times 40$, $60\times 60$, $80\times 80$, and $100\times 100$ GEMM (PE with Load-Store CFU, with DOT instruction (PE-Architectural Enhancement $2$ (AE2))\label{tab:mm3}}{
\begin{tabular}{ |c|c|c|c|c|c| } 
 \hline
 Matrix Size & $20\times 20$ & $40\times 40$ & $60\times 60$ & $80\times 80$ & $100\times 100$ \\  \hline \hline
 Latency (in clock cycles) & 15251 & 113114 & 371699 & 877124 & 1696921 \\ \hline 
Improvement over table \ref{tab:mm2} & 33.7\% & 36.6\% & 37.57\% & 37.82\% & 37.85\% \\ \hline
Gflops/watt  & 10.52 & 11.49 & 11.85 & 11.93 & 12.06 \\ \hline
\end{tabular}}
\end{center}
\end{table*} 

It can be observed from the table \ref{tab:mm3} that as we increase the matrix size, the benefit due to DOT instruction improves. This is due to improved exploitation of ILP in the FPS.

	\subsubsection{Block Data Load and Block Data Store Instructions}
We further aim to reduce handshaking between LM and GM. This reduction in the handshaking in-turn improves data availability in the Register File. 
In order to reduce the handshaking between PE and the next level of the Memory, we introduce instructions that can load/store data in a block fashion. Performance improvement due to Block Data Load and Block Data Store is shown in table \ref{tab:mm4} where we have used $4\times 4$ as a block size for the transfer. 

\begin{table*}[]
\begin{center}
\caption{Latencies of $20\times 20$, $40\times 40$, $60\times 60$, $80\times 80$, and $100\times 100$ GEMM (PE with Load-Store CFU, with DOT4, and Block Data Load/Store instructions (PE-Architectural Enhancement $3$ (AE3))\label{tab:mm4}}{
\begin{tabular}{ |c|c|c|c|c|c| } 
 \hline
 Matrix Size & $20\times 20$ & $40\times 40$ & $60\times 60$ & $80\times 80$ & $100\times 100$ \\  \hline \hline
 Latency (in clock cycles) & 12745 & 97136 & 324997 & 784838 & 1519083\\ \hline 
Improvement over table \ref{tab:mm3} & 16.4\% & 14.1\% & 12.5\% & 10.51\% & 10.48\% \\ \hline
Gflops/watt  & 12.59 & 13.38 & 13.56 & 13.33 & 13.47 \\ \hline
\end{tabular}}
\end{center}
\end{table*} 

It can be observed from the table \ref{tab:mm4} that as we increase matrix size, the benefit due to Block Data Load/Store does not improve. Rather the performance is observed to be saturating. This is because of the constant block size of $4\times 4$ across all the matrix sizes. Supporting larger block size is not possible due to limited registers availability in the Register File in the FPS. It can also be observed in table \ref{tab:mm4} that the latency gap between $20\times 20$ and $40\times 40$, and $40\times 40$ and $60\times 60$ is also decreasing and it is likely to saturate at some point. Further experiments show that the gap saturates at $10\%$ for larger matrix sizes.

\subsection{Bandwidth Increase}
Increased resources in the FPS improves performance by almost $2$x, and reduced handshaking between LM and the upper level of the memory improves performance by 10\%. We still see significant gap between our desired performance and attained performance.  The reason for this gap is mainly because of under utilization of the RDP that is configured as DOT4. In order to improve resource utilization of RDP, in this architectural enhancement, we increase bandwidth between FPS and Load-Store CFU to $4$ times. We consider increase in the bandwidth to $4$ times since the block size supported in FPS is $4\times 4$. We transfer $256$-bits between FPS and Load-Store CFU in contrast to previous realization where we transfered $64$-bit data. The communication between FPS and Load-Store CFU at higher rate ensures better data availability in the Register File of FPS. The performance improvement due to increase in the bandwidth is shown in table \ref{tab:mm5}.

\begin{table*}[]
\begin{center}
\caption{Latencies of $20\times 20$, $40\times 40$, $60\times 60$, $80\times 80$, and $100\times 100$ GEMM (PE with Load-Store CFU, with DOT and Block Data Load/Store instructions, increased bandwidth (PE-Architectural Enhancement $4$ (AE4))\label{tab:mm5}}{
\begin{tabular}{ |c|c|c|c|c|c| } 
 \hline
 Matrix Size & $20\times 20$ & $40\times 40$ & $60\times 60$ & $80\times 80$ & $100\times 100$ \\  \hline \hline
 Latency (in clock cycles) & 7079 & 52624 & 174969 & 422924 & 818178 \\ \hline 
Improvement over table \ref{tab:mm4} & 44.4\% & 45.8\% & 46.1\% & 46.12\% & 46.14\% \\ \hline
Gflops/watt  & 22.67 & 24.71 & 25.19 & 24.95 & 25.02 \\ \hline
\end{tabular}}
\end{center}
\end{table*}

It can be observed in table \ref{tab:mm5} that as we increase the matrix size the benefits due to increased bandwidth between FPS and Load-Store CFU in the PE improves. This is mainly because of better utilization of RDP (here configured as DOT4).

\subsection{Pre-fetching}
To improve the utilization of RDP further in the FPS, we restructure the loop in GEMM. We re-write algorithm \ref{algo:gmm1} as algorithm \ref{algo:gmm_prefetch}.

\begin{algorithm}
\caption{General Matrix Multiplication with Pre-fetching}
\label{algo:gmm_prefetch}
\begin{algorithmic}[1]
\State Allocate memories for input and output matrices
\For{$i=1$ to $m$}
  \For{$j=1$ to $n$}
     \State $C[i][j] = A[i][k]\times B[k][j]$ 
    \For{$k=1$ to $n$}
      \State $C[i][j] = A[i][k]\times B[k][j]  + C[i][j]$
     \EndFor
  \EndFor
\EndFor
\end{algorithmic}
\end{algorithm}


\begin{figure}[]
	\centering
	\includegraphics[scale=0.13]{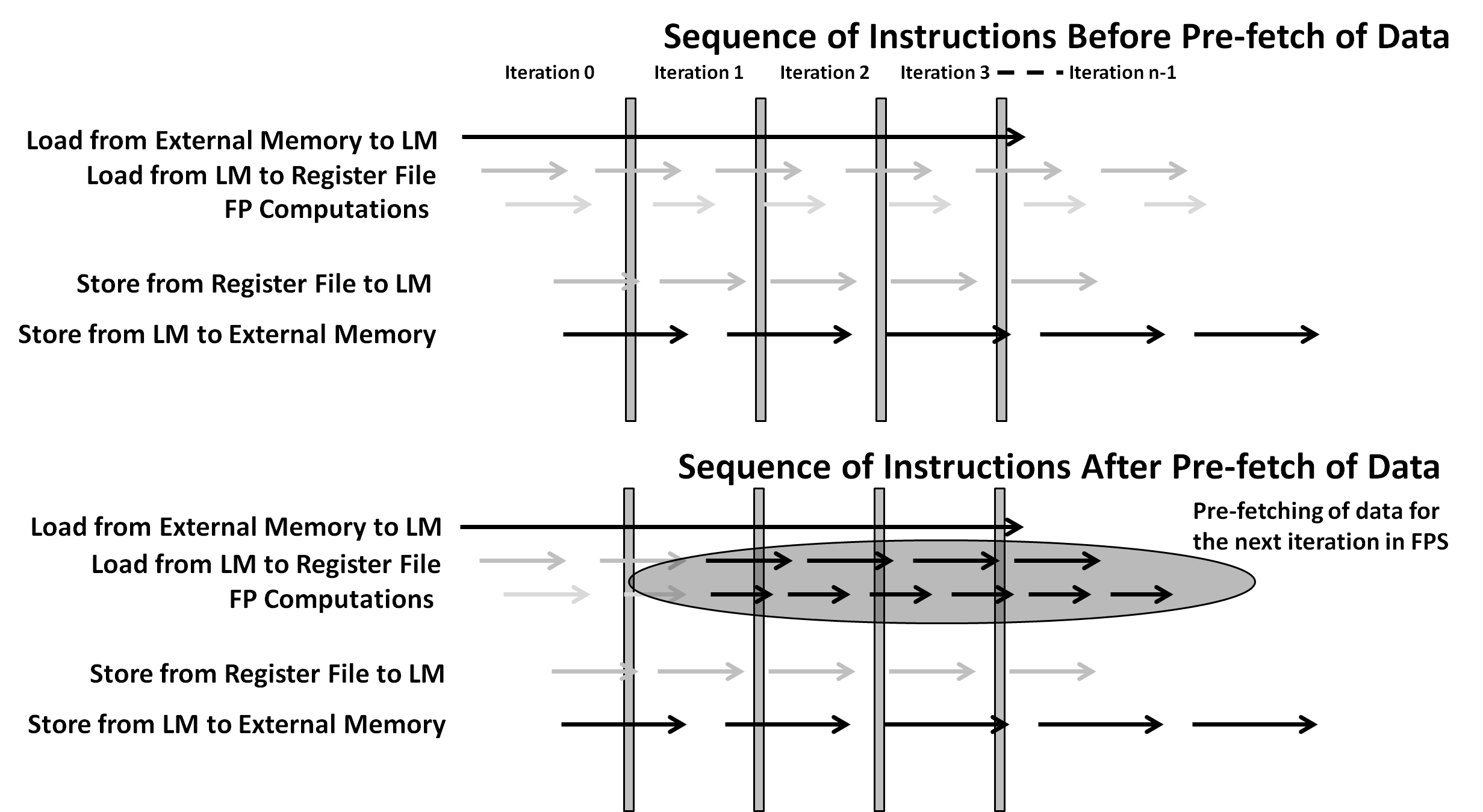}
	\caption{Pre-fetching Matrix Elements for the Next Iteration \cite{Merc1}}
	\label{fig:st}
\end{figure}

Re-structuring the loops in the algorithm allows us to pre-fetch the matrix block required for the next iteration. This results in better exploitation of FPU pipeline by reduced instruction stalls in FPS as shown in figure \ref{fig:st}. The shaded portion in the figure \ref{fig:st} depicts the reduction in the instruction stalls in FPS when there is a pre-fetch of the block of the matrix required in the next iteration for computation. In figure \ref{fig:st}, there are two portions, 1) before pre-fetching, and 2) after pre-fetching. Arrows in the figures depict execution of different types of operations such as computations in FPS, loading/storing of data from/to GM (or EM) memory, loading/storing of data from/to GM.   

\begin{table*}[]
\begin{center}
\caption{Latencies of $20\times 20$, $40\times 40$, $60\times 60$, $80\times 80$, and $100\times 100$ GEMM (PE with Load-Store CFU, with DOT and Block Data Load/Store instructions, increased bandwidth and data pre-fetching (PE-Architectural Enhancement $5$ (AE5))\label{tab:mm6}}{
\vspace{-3mm}
\begin{tabular}{ |c|c|c|c|c|c| } 
 \hline
 Matrix Size & $20\times 20$ & $40\times 40$ & $60\times 60$ & $80\times 80$ & $100\times 100$ \\  \hline \hline
 Latency (in clock cycles) & 5561 & 38376 & 124741 & 298161 & 573442 \\ \hline 
Improvement over table \ref{tab:mm5} & 21.44\% & 27.07\% & 28.70\% & 29.5\% & 29.9\%\\ \hline
Gflops/watt & 28.86 & 33.88 & 35.33 & 35.11 & 35.70 \\ \hline
\end{tabular}}
\end{center}
\end{table*}

Improvement attained by pre-fetching is shown in table \ref{tab:mm6}. It can be observed in the table \ref{tab:mm6} that as we increase matrix size the benefits due to pre-fetching increases. This is mainly because of improvement in data availability in the Register File of the FPS.

\begin{figure*}[]
\centering
\subfigure[Reduction in the Latency in DGEMM Due to Architecture Enhancements\label{fig:mm_explo1}]{\includegraphics[scale = 0.17]{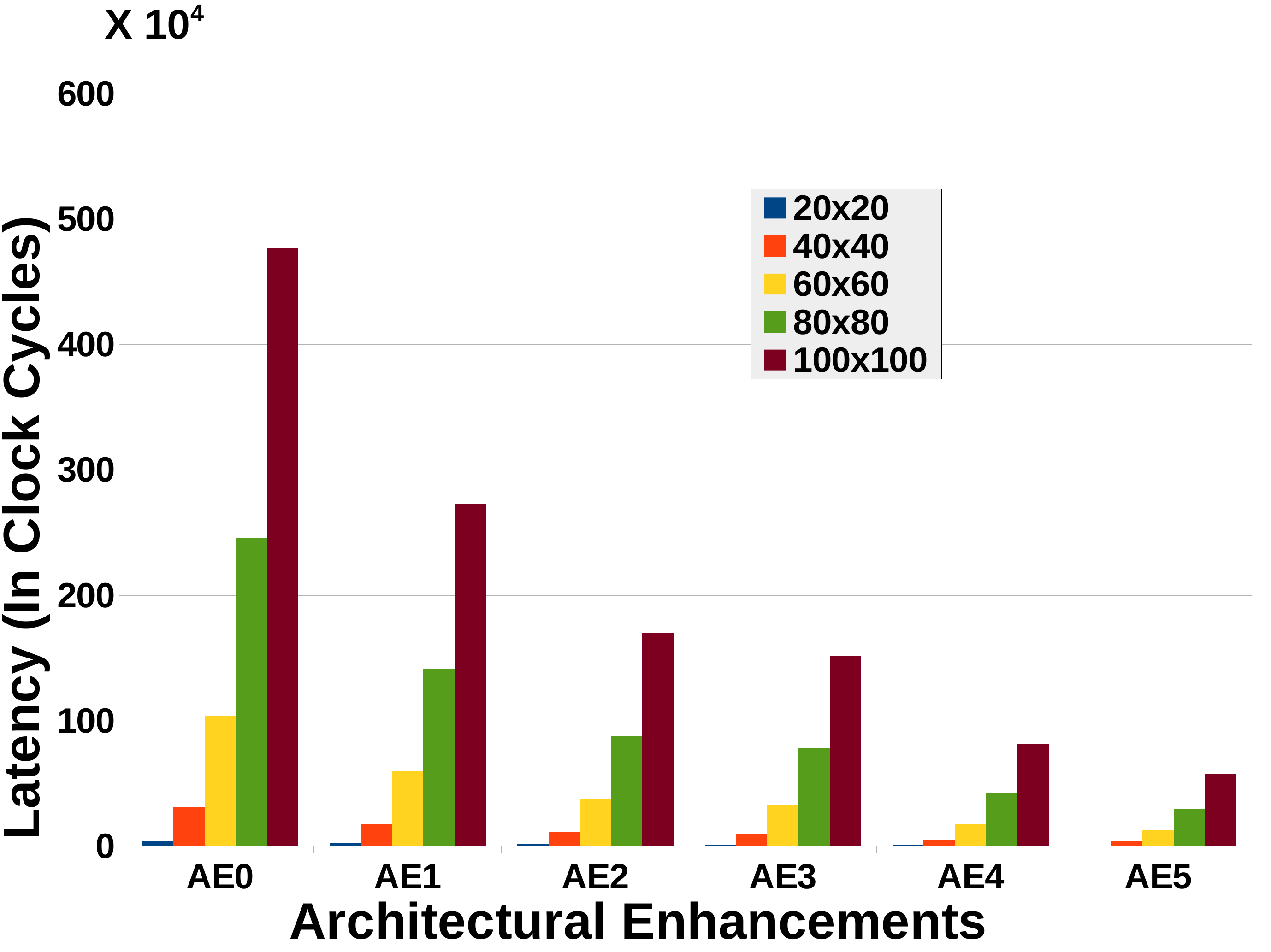}}
\subfigure[Latencies Normalized to Total Computations in terms of DOT4 in DGEMM\label{fig:mm_explo2}]{\includegraphics[scale = 0.17]{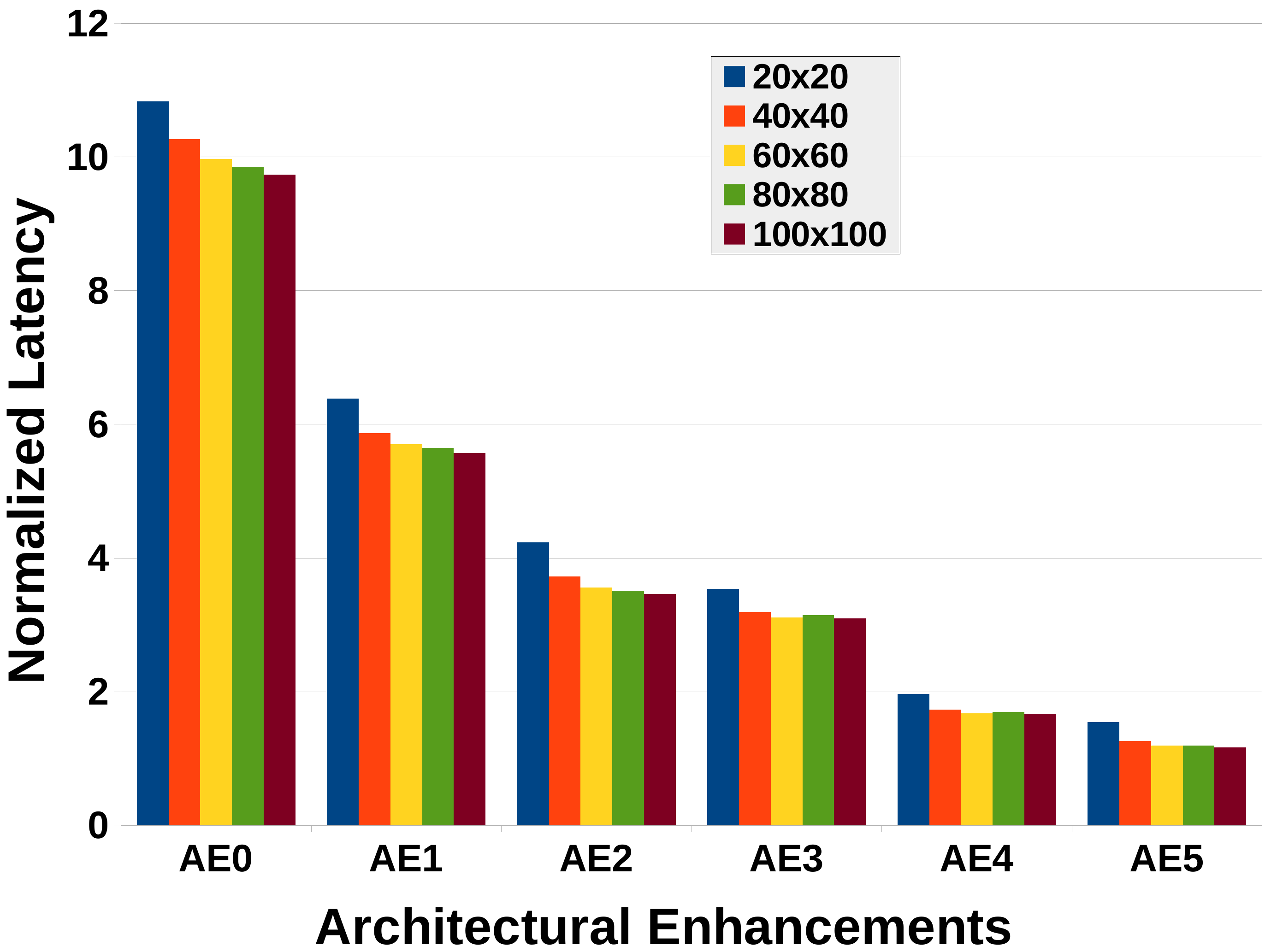}}
\subfigure[Cycles per Floating Point Operation in DGEMM \label{fig:mm_explo3}]{\includegraphics[scale = 0.17]{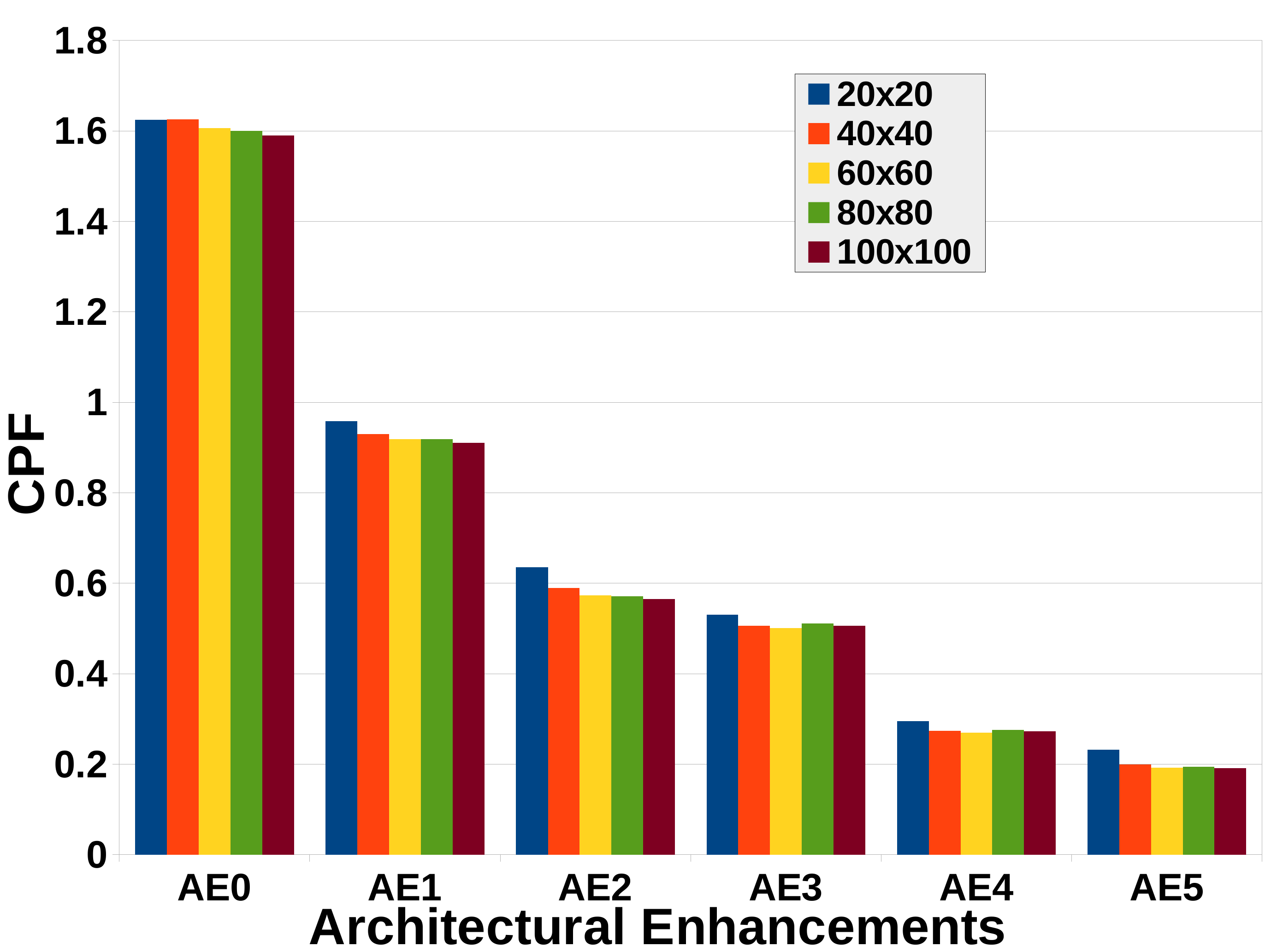}}
\subfigure[Floating Point Operations per Cycle in DGEMM \label{fig:mm_explo4}]{\includegraphics[scale = 0.17]{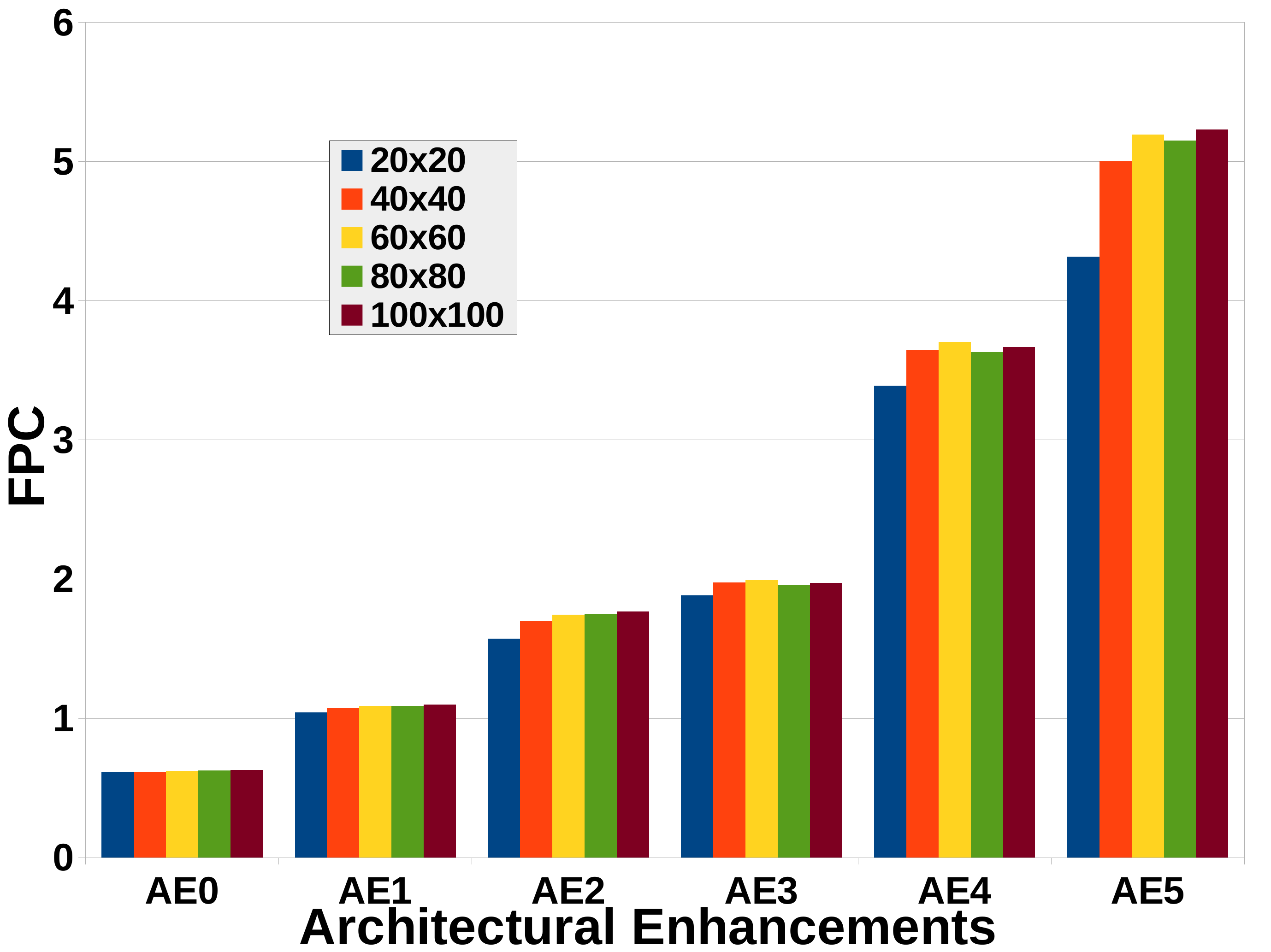}}
\subfigure[Percentage of Peak FPC in DGEMM with Each Architectural Enhancement\label{fig:mm_explo5}]{\includegraphics[scale = 0.17]{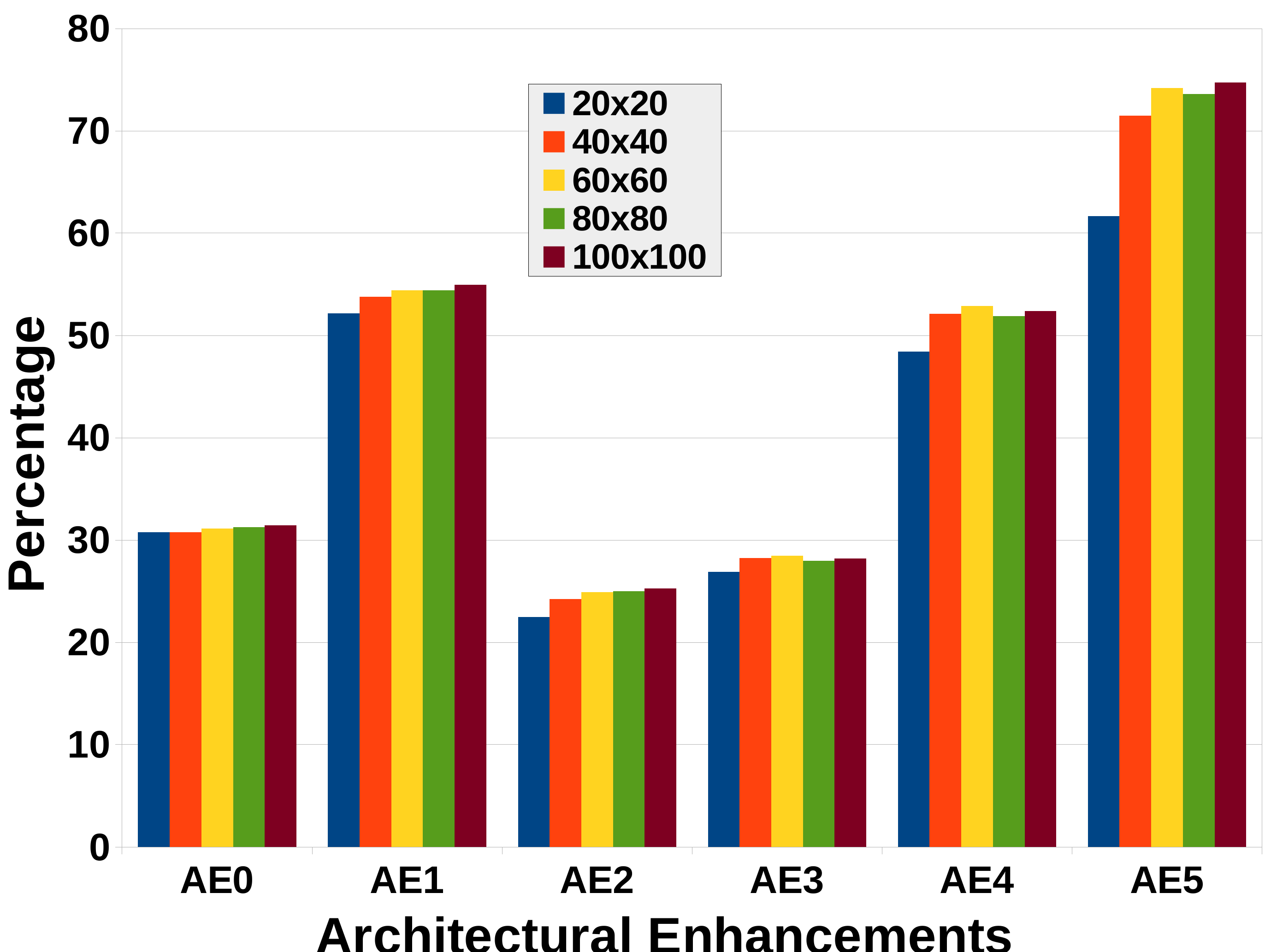}}
\subfigure[Percentage of Peak FPC in DGEMV with Each Architectural Enhancement\label{fig:mm_explo6}]{\includegraphics[scale = 0.17]{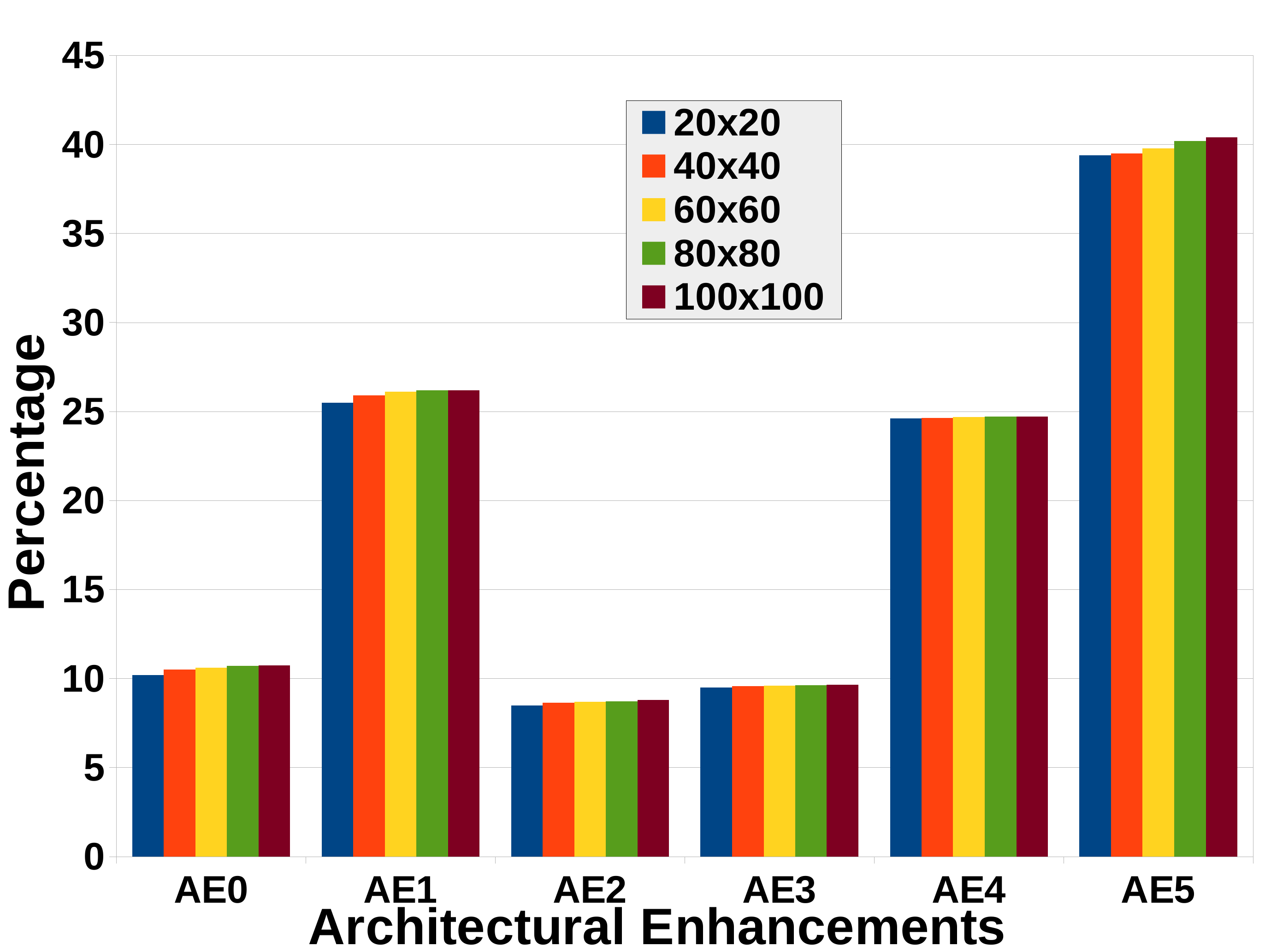}}
\subfigure[Percentage of Peak FPC in DDOT with Each Architectural Enhancement\label{fig:mm_explo7}]{\includegraphics[scale = 0.17]{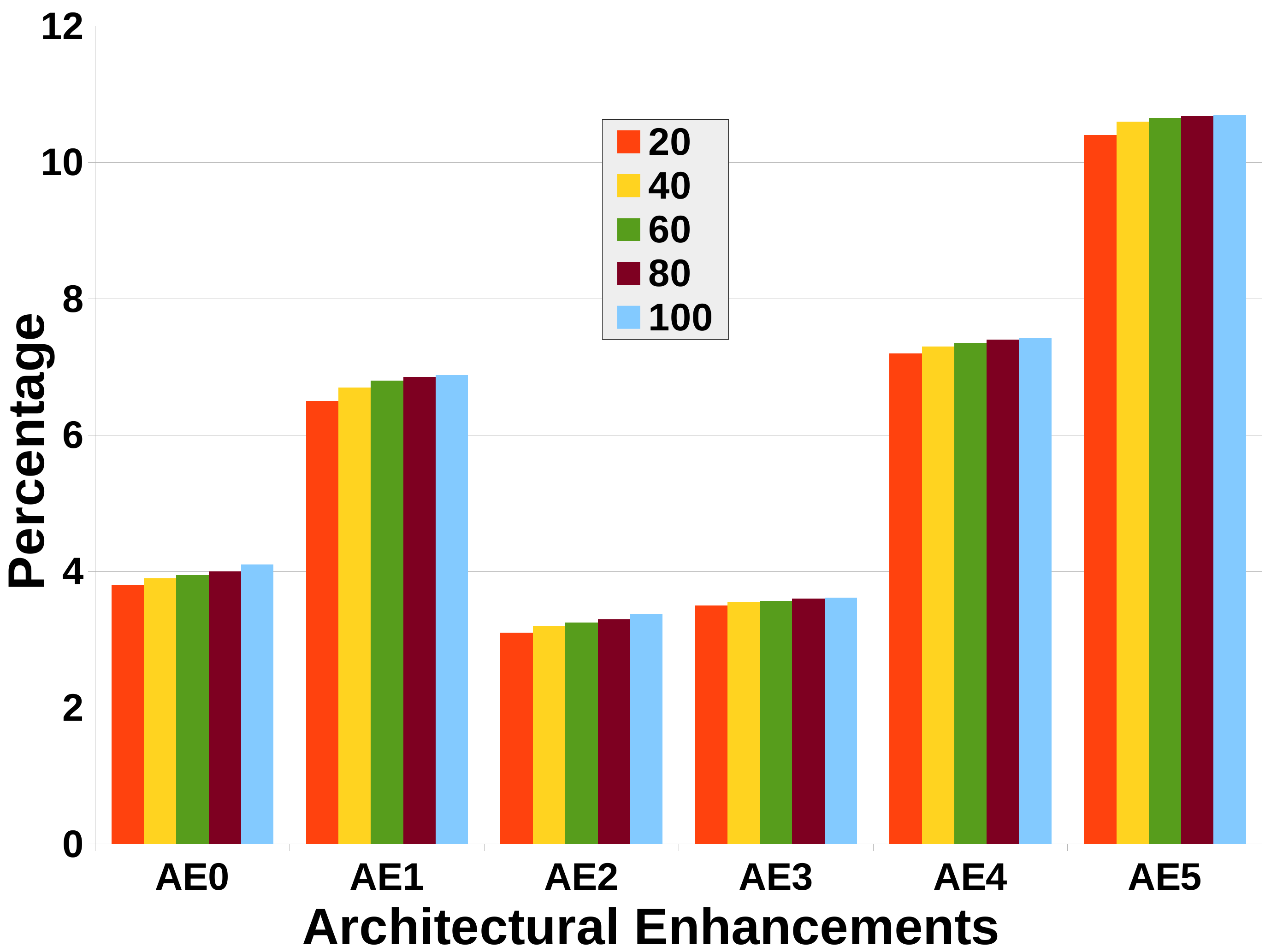}}
\subfigure[Gflops/watt for DGEMM at 0.2GHz, 0.33GHz, 0.95GHz, and 1.81GHz\label{fig:mm_explo8}]{\includegraphics[scale = 0.17]{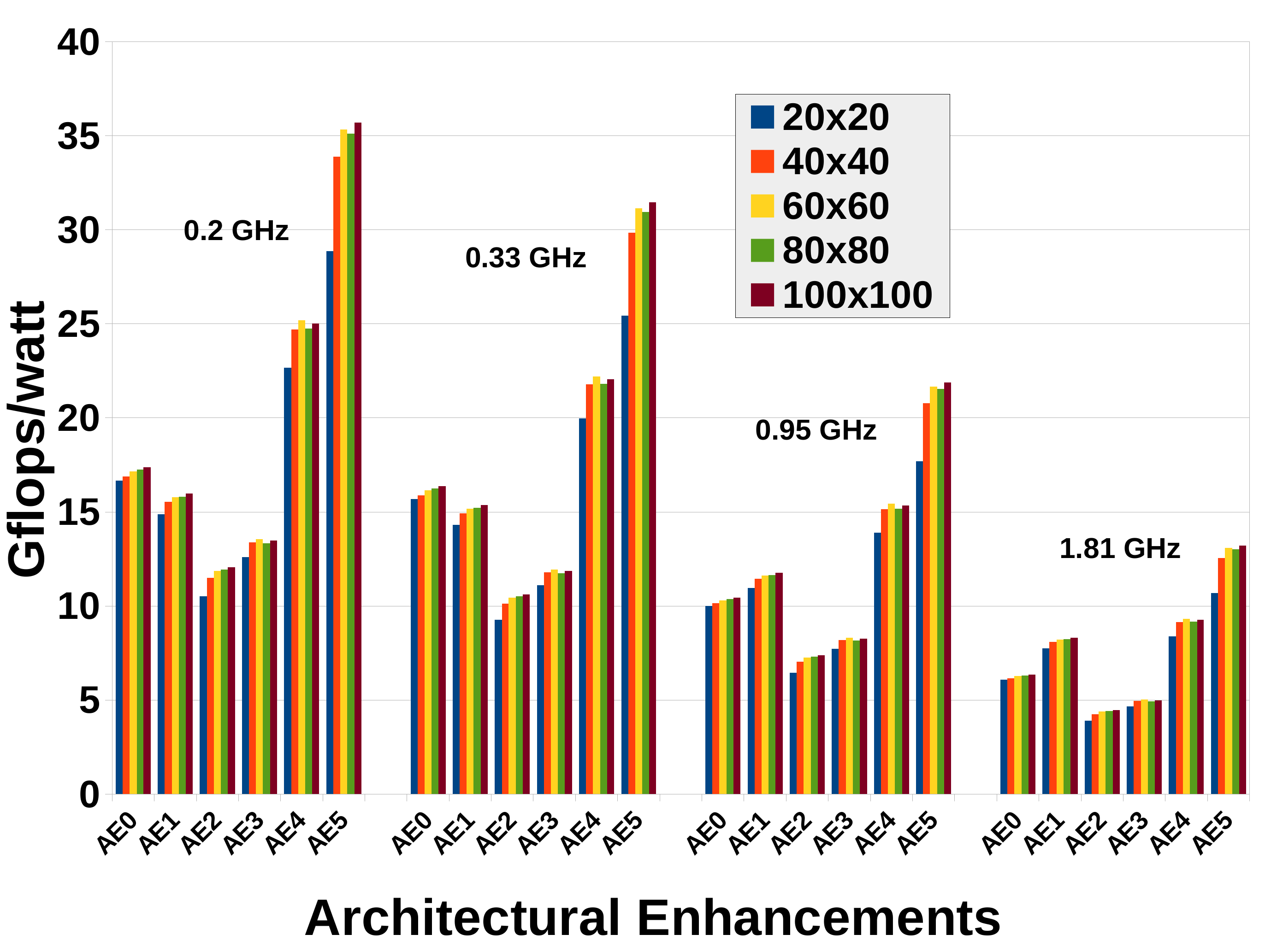}}
\subfigure[Gflops/mm$^2$ for DGEMM at 0.2GHz, 0.33GHz, 0.95GHz, and 1.81GHz \label{fig:mm_explo9}]{\includegraphics[scale = 0.17]{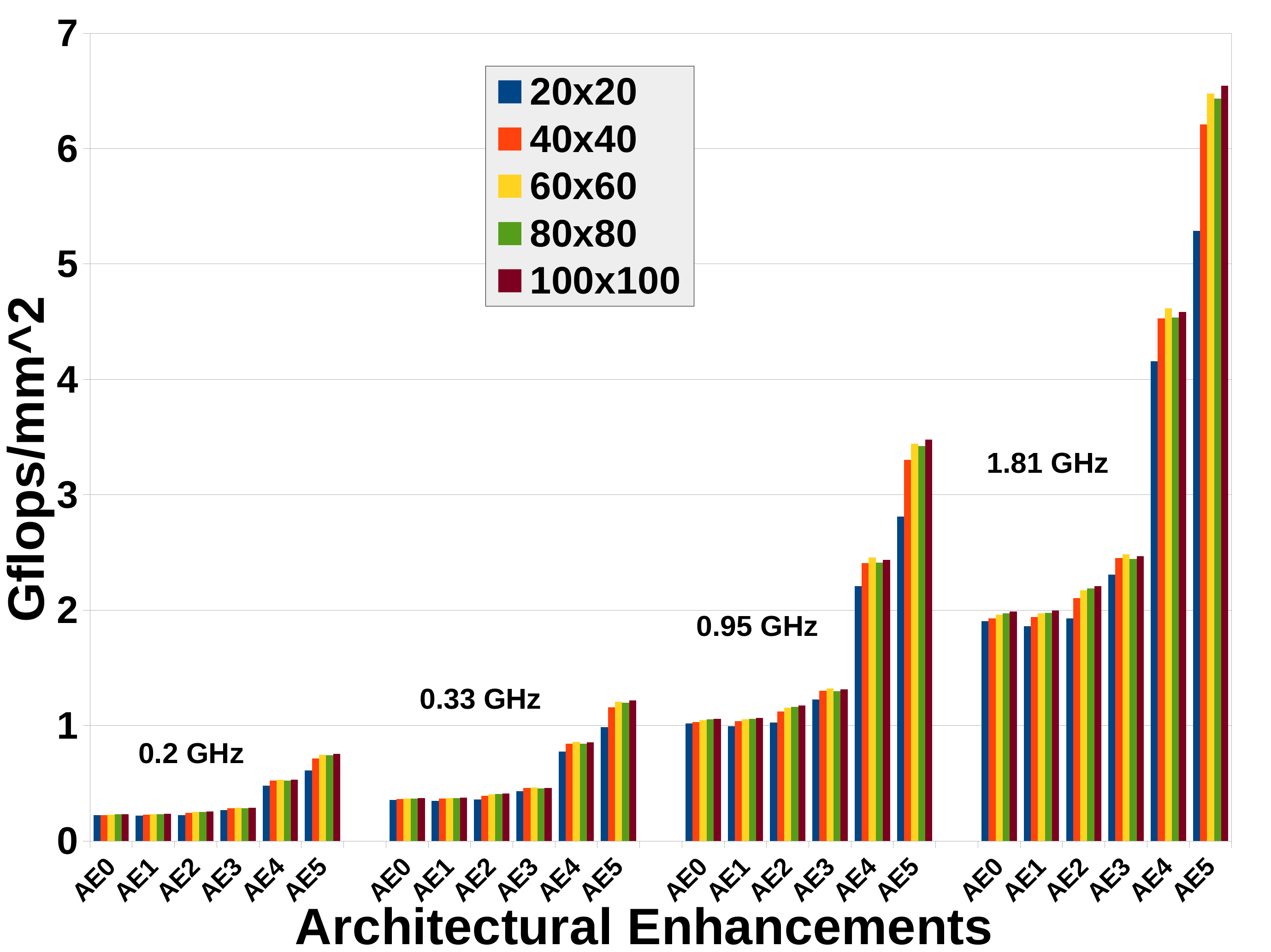}}
\subfigure[Performance Comparison of REDEFINE-PE with Other Platforms\label{fig:mm_explo10}]{\includegraphics[scale = 0.17]{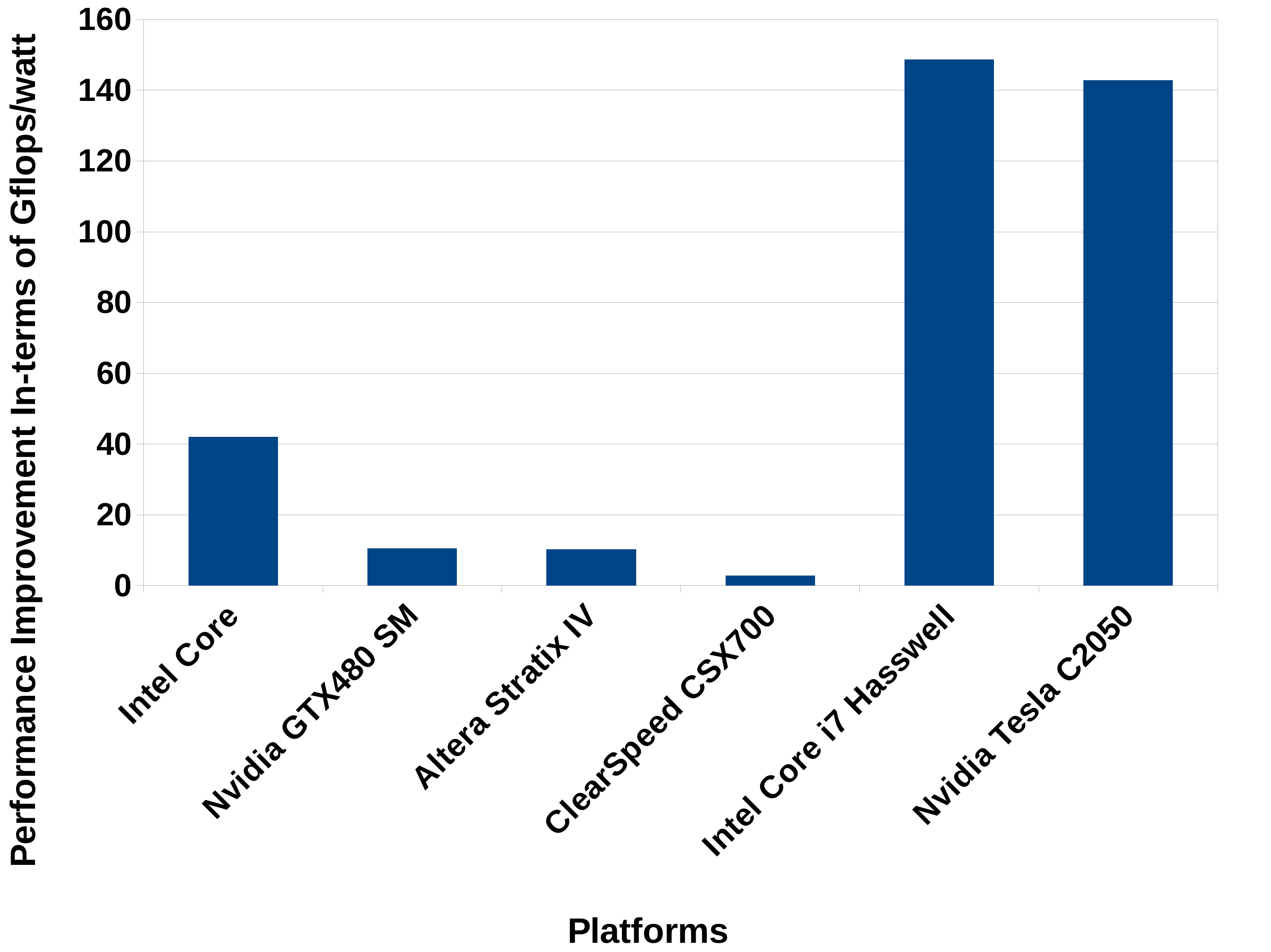}}
\subfigure[Simulation Environment where Tile array of $2\times 2$ is used for realizing DGEMM\label{fig:mm_2x2}]{\includegraphics[scale = 0.17]{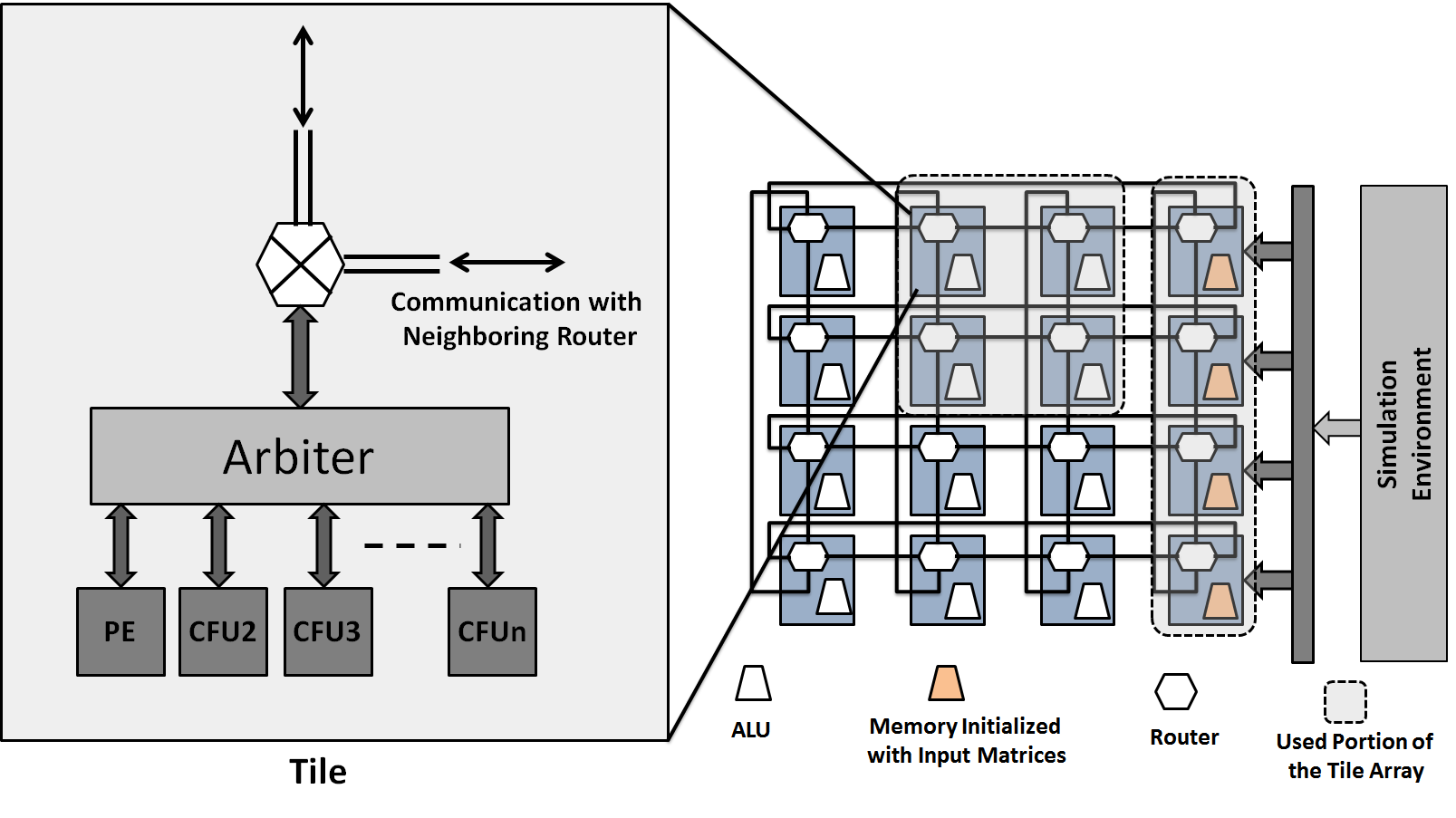}}
\caption{Performance of DGEMM}
\label{fig:dgemm_perf}
\end{figure*}



Collectively, the reduction in execution cycles of GEMM can be seen in the figure \ref{fig:mm_explo1} as we perform different architectural enhancements in the PE. It can be clearly observed that finally we get speed-up of 7x for matrix of size $20\times 20$, 8.13x for the matrix of size $40\times 40$, and 8.34x for the matrix of size $60\times 60$ \cite{Merc1}. 

It can be observed from figure \ref{fig:mm_explo2} that as we perform different architectural enhancements, the ratio of Latency to computations reduces. If we denote the ratio of Latency to total computations by $\alpha$ then  

\begin{align}
	\alpha = \frac{Latency}{Total\;Computations\; in\;Terms\;of\;DOT4}
\end{align}

It can be observed in the figure \ref{fig:mm_explo2} that as we increase matrix size $\alpha$ asymptotically approaches $1$. $\alpha = 1$ is a case where there is a complete overlap of computation and communication. Complete overlap of computations and communication is not possible in the real life scenario and hence $\alpha$ can never become $1$.

Figure \ref{fig:mm_explo3} depicts CPF for matrices of size $20\times 20$, $40\times 40$, and $60\times 60$. It can be observed in the figure \ref{fig:mm_explo3} that as we perform enhancements the CPF tends to decrease and this trend is observed across all the matrices. Figure \ref{fig:mm_explo4} depicts FPC (where FPC = 1/CPF). It can be observed from figure \ref{fig:mm_explo4} that, as we perform enhancements in the PE, FPC improves dramatically. Although CPF and FPC are good measure of performance of a PE, they do not convey enough information about how efficiently compute resources are utilized in the PE. 

Percentage of peak FPC attained in PE after every enhancement is shown in figure \ref{fig:mm_explo5}. Figure \ref{fig:mm_explo5} depicts an interesting trend where the peak FPC reduces drastically and then with further architectural enhancements, improves. As our enhancements suggests, in the first enhancement where we place Load-Store CFU with LM for overlap of computation and communication, the FPC achieved saturates at 54\% of the peak FPC\footnote{Here peak FPC = $\frac{1}{CPF} = \frac{1}{\frac{1}{2}} = 2$}. We strongly intend to break this saturation point since 54\% of the peak is not a satisfactory performance. In order to break this saturation point we further enhance FPS with compute resources that leads to higher theoretical peak FPC\footnote{Here peak FPC = $\frac{1}{CPF} = \frac{1}{\frac{1}{7}} = 7$. Increase in peak FPC is due to DOT4 instruction}. At this point achieved FPC reduces at $AE2$ as shown in the figure \ref{fig:mm_explo5}. This is because of increased compute resources. Our further enhancements help us to improve the resource utilization of the increased resources in FPS and achieve up-to 74\% of the peak FPC of the PE. 

We presented methodical architectural enhancements to improve the performance of PE. Through architectural customizations, we could break saturation point at 54\% and improve performance of PE. In other words, we showed that the performance of the algorithms can be improved by customizations that are specific to the algorithms. Here, we showed this with example of GEMM which is a Level-3 BLAS. Finally, 35.7 Gflops/watt in the PE is achieved through carefull realization of Level-3 BLAS.      

\subsection{Parallel Realization of Level-3 BLAS}

For parallel realization of DGEMM, our two different simulation environments are shown in figure \ref{fig:mm_2x2}. In figure \ref{fig:mm_2x2}, shaded portion of the Tile array (except the last column) which is $2\times 2$ Tiles is used for the computations where we realize DGEMM while the last column is used for storing input and output matrices. We use Octave for generating input matrices. Similarly, we use $3\times 3$ portion of the Tile array for realizing DGEMM. 


In our experiments, if output matrix is of size $n\times n$ then we divide the output matrix into blocks of $\frac{n}{b}\times \frac{n}{b}$ where $b\times b$ is the Tile array that we are using. In our experiments $b=2$ or $3$. For example, if output matrix is of size $20\times 20$, and we are using $2\times 2$ of the Tile array to compute the DGEMM, we divide output matrices into $10\times 10$ block matrices. Now, in each Tile that we are using, we compute one of the block of size $10\times 10$. Similarly, if output matrix is of size $60\times 60$, and Tile array that is used for computing DGEMM is of size $3\times 3$ then block of $20\times 20$ is computed in each of the Tile of REDEFINE.

\begin{figure}[!ht]
	\centering
	\includegraphics[scale = 0.20]{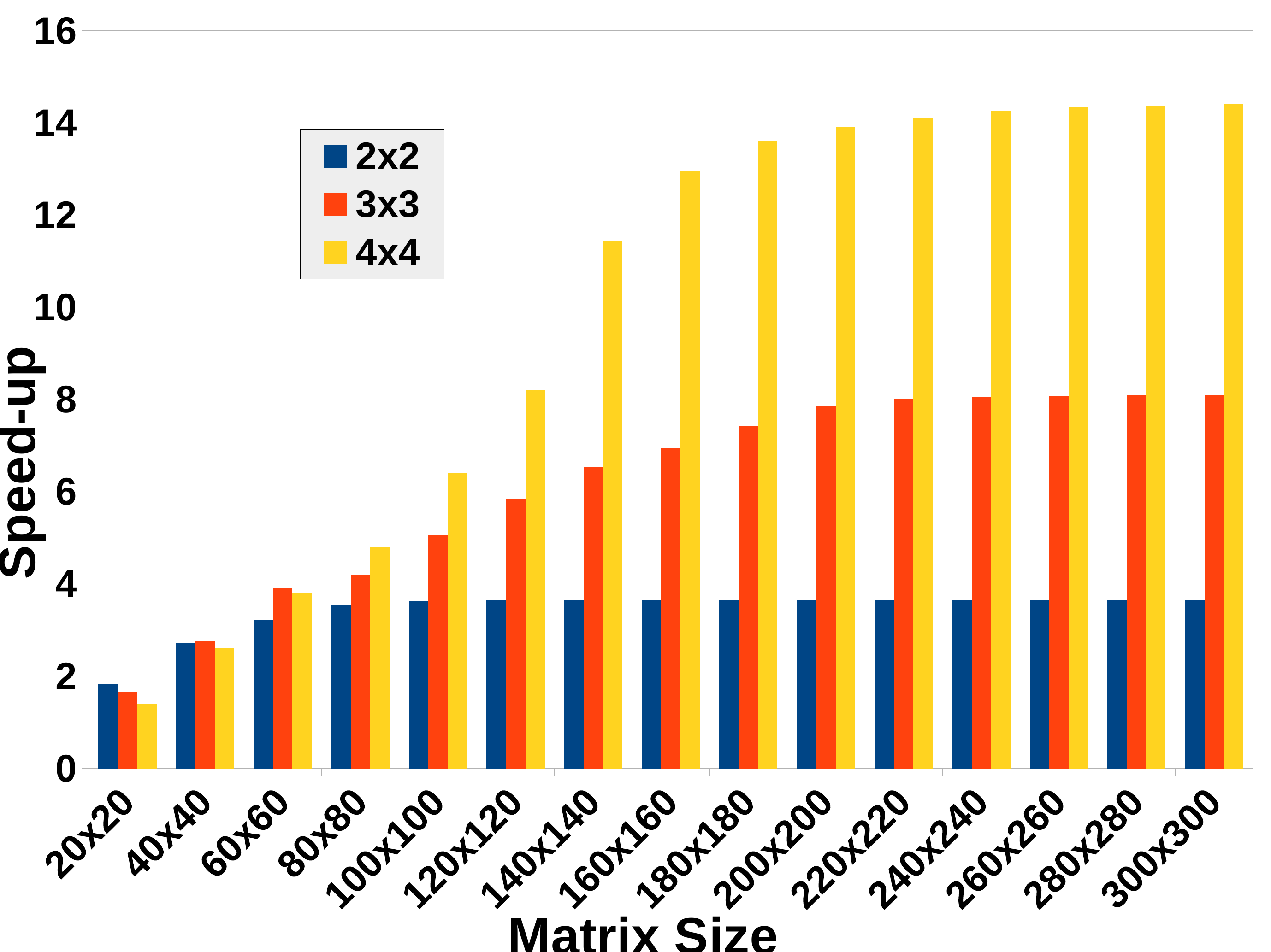}
	\caption{Speed-up in REDEFINE for DGEMM for Different Configurations}
	\label{fig:final_res1}
\end{figure}


Speed-up attained in REDEFINE is is shown in the figure \ref{fig:final_res1} when Tile array of size $2\times 2$, $3\times 3$, and $4\times 4$ are used for the experiments. It can be observed from the figure \ref{fig:final_res1} that when we use Tile array of $2\times 2$ the speed-up over PE realization approaches $4$ as we increase matrix size. When we use Tile array of $3\times 3$ the speed-up over PE realization approaches $9$, and for Tile array of size $4\times 4$ the speed-up attained approaches 16. For small matrices, communication with the last column of the Tile array is dominant over computations in the Tile. For example, for a matrix of size $20\times 20$ and Tile array size of $2\times 2$, each Tile computes $10\times 10$ block of the resultant matrix. Ignoring the coefficient of the highest order term, there will be $10^3$ computations over $10^2$ loads/stores. Computation to communication ratio in for $20\times 20$ matrix will be $10$ in each Tile. For matrix a of size  $60\times 60$ where each Tile will compute a block of $20\times 20$ matrix, computation to communication ratio is $20$. One more observation we make here is that, as we increase the matrix size, speed-up in REDEFINE over PE saturates. This is because of the saturation in the parallelism exploited by the PE that is attached in each Tile of REDEFINE.

In this section, we presented results for the PE that we presented in section \ref{sec:pe_design}. We use the estimation methodology presented in \cite{lac1}, \cite{lac2}, and \cite{lac3} for fair comparison of the platforms. As shown in the figure \ref{fig:mm_explo10}, it can be observed that the performance of the PE is 40-140x better than Intel Core architectures while 7-139x better than Nvidia GPUs. Compared to Altera FPGA, PE is 10x better in terms of Gflops/watt while compared to ClearSpeed CSX700 it is almost 3x better.

\section{Conclusion}\label{sec:con}
While the recent realizations for matrix computations focus on architectural customization for DLA, in this paper we presented a novel way of algorithm-architecture co-design for breaking the performance saturation point in BLAS and presented a systematic enhancements in the micro-architecture for exploiting underlying compute and memory resources more efficiently. In algorithm-architecture co-design, we first realized Level-3 BLAS on off-the-shelf processors and exhibited that the performance achieved by these off-the-shelf processors in DGEMM is not satisfactory. The performance in the off-the-shelf Intel and AMD processors saturates at 15-17\% at 65W and 57\% in Nvidia Tesla C2050 GPGPU with the best optimization efforts. This dis-satisfactory performance is mainly due to inefficiently exploited compute and memory resources of the underlying platform. We could sense a scope here in breaking this performance saturation point in a custom architecture and performed analysis of BLAS, and designed a PE that could efficiently execute BLAS routines at much higher Gflops/watt. We further enhanced this PE with several features such that we could efficiently exploit compute resources and memory resources in the PE and achieve performance of 35.7 Gflops/watt that is much higher than the off-the-shelf Intel and AMD processors and GPGPUs. We attached this PE in REDEFINE for parallel realization of BLAS and showed that the speed-up achieved in the parallel realization is commensurate with the number of Tiles used and hence we showed that our solution is scalable. 

\bibliographystyle{IEEEtran}
\bibliography{IEEEabrv,ref}

\begin{IEEEbiography}[{\includegraphics[width=1in,height=1.25in,clip,keepaspectratio]{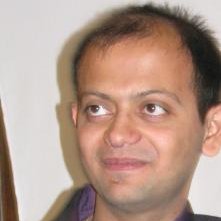}}]{Farhad Merchant}
 Farhad Merchant is a Research Fellow at Hardware and Embedded Systems Lab, School of Computer Science and Engineering, Nanyang Technological University, Singapore. He received his PhD from Computer Aided Design Laboratory, Indian Institute of Science, Bangalore, India. His research interests are algorithm-architecture co-design, computer architecture, reconfigurable computing, development and tuning of high performance software packages
\end{IEEEbiography}
\vspace{-10mm}

\begin{IEEEbiography}[{\includegraphics[width=1in,height=1.25in,clip,keepaspectratio]{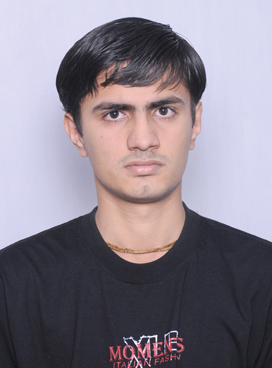}}]{Tarun Vatwani}
Tarun Vatwani is a fresh B.Tech. graduate from Indian Institute of Technology, Jodhpur, India, His research interests are computer architecture, high performance computing, machine learning, performance tuning of different software packages.
\end{IEEEbiography}
\vspace{-10mm}

\begin{IEEEbiography}[{\includegraphics[width=1in,height=1.25in,clip,keepaspectratio]{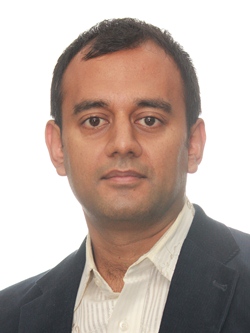}}]{Anupam Chattopadhyay}
 Anupam Chattopadhyay received his B.E. degree from Jadavpur University, India in 2000. He received his MSc. from ALaRI, Switzerland and PhD from RWTH Aachen in 2002 and 2008 respectively. From 2008 to 2009, he worked as a Member of Consulting Staff in CoWare R\&D, Noida, India. From 2010 to 2014, he led the MPSoC Architectures Research Group in RWTH Aachen, Germany as a Junior Professor. Since September, 2014, he is appointed as an assistant Professor in SCE, NTU.
\end{IEEEbiography}
\vspace{-10mm}
\begin{IEEEbiography}[{\includegraphics[width=1in,height=1.25in,clip,keepaspectratio]{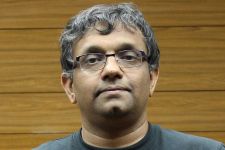}}]{Soumyendu Raha}
 Soumyendu Raha obtained his PhD in Scientific Computation from the University of Minnesota in 2000.
Currently he is a Professor of the Computational and Data Sciences Department at the Indian
Institute of Science in Bangalore, which he joined in 2003, after having worked for IBM for a couple of years.
His research interests are in computational mathematics of dynamical systems, both continuous and combinatorial,
and in co-development and application of computing systems for implementation of computational mathematics algorithms.
\end{IEEEbiography}
\vspace{-10mm}

\begin{IEEEbiography}[{\includegraphics[width=1in,height=1.25in,clip,keepaspectratio]{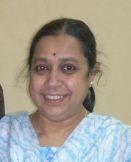}}]
{Ranjani Narayan} Dr. Ranjani Narayan has over 15 years experience at
IISc and 9 years at Hewlett Packard. She has vast work
experience in a variety of fields – computer architecture,
operating systems, and special purpose systems. She
has also worked in the Technical University of Delft, The
Netherlands, and Massachusetts Institute of Technol-
ogy, Cambridge, USA. During her tenure at HP, she
worked on various areas in operating systems and
hardware monitoring and diagnostics systems. She has
numerous research publications.She is currently Chief
Technology Officer at Morphing Machines Pvt. Ltd,
Bangalore, India.
\end{IEEEbiography}
\vspace{-10mm}

\begin{IEEEbiography}[{\includegraphics[width=1in,height=1.25in,clip,keepaspectratio]{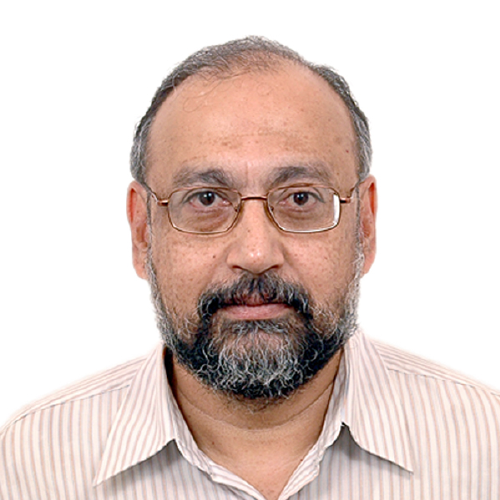}}]{S K Nandy}
S. K. Nandy is a Professor in the Department of Computational and Data Sciences of the Indian Institute of Science, Bangalore.
His research interests are in areas of High Performance Embedded Systems on a  Chip, VLSI architectures for Reconfigurable Systems on Chip,
and Architectures and Compiling Techniques for Heterogeneous Many Core Systems. Nandy received the B.Sc (Hons.) Physics degree from the Indian Institute of Technology, Kharagpur, India, in 1977. He obtained the BE (Hons.) degree
 in Electronics and Communication in 1980, MSc.(Engg.) degree in Computer Science and Engineering in 1986, and the Ph.D. degree in Computer Science
and Engineering in 1989 from the Indian Institute of Science, Bangalore. He has over 170 publications in International Journals, and Proceedings of International Conferences, and 5 patents.
\end{IEEEbiography}

\end{document}